\documentclass[aps,pra,twocolumn,superscriptaddress,floatfix,footinbib]{revtex4-2}
\usepackage[english]{babel}
\usepackage{graphicx}
\usepackage{float}
\usepackage{lipsum}
\usepackage{amsmath}
\usepackage{braket}
\usepackage{subcaption}
\usepackage{placeins}
\usepackage{xspace}
\usepackage{siunitx}

\makeatletter
\DeclareRobustCommand{\au}[1]{%
    \@ifnextchar{.}
    {\SI{#1}{{a.u}}}
    {\SI{#1}{{a.u.}}\ifmmode\else\@\xspace\fi}%
}
\makeatother

\begin{document}

\title{Excited $\Sigma$ states of the hydrogen-antihydrogen molecule}
\author{L.~Brumm}
\affiliation{AG Moderne Optik, Institut f\"{u}r Physik, Humboldt-Universit\"{a}t zu Berlin, Newtonstr.~15, 12489 Berlin, Germany}
\author{J.~Sch\"urmann}
\affiliation{AG Moderne Optik, Institut f\"{u}r Physik, Humboldt-Universit\"{a}t zu Berlin, Newtonstr.~15, 12489 Berlin, Germany}
\affiliation{Institute for Nuclear Physics, Westf\"{a}lische Wilhelms-Universit\"{a}t M\"{u}nster, Wilhelm-Klemm-Str.~9, 48149 M\"{u}nster, Germany}
\author{A.~Saenz}
\affiliation{AG Moderne Optik, Institut f\"{u}r Physik, Humboldt-Universit\"{a}t zu Berlin, Newtonstr.~15, 12489 Berlin, Germany}

% Command Section
\newcommand{\Hbar}{$\bar{\mathrm{H}}$}
\newcommand{\HHbar}{$\mathrm{H}\bar{\mathrm{H}}$}
\newcommand{\KW}{Ko\l{}os-Wolniewicz}
\newcommand{\KaW}{W.~Ko\l{}os and L.~Wolniewicz}
\newcommand{\rpe}{r_{pe}}
\newcommand{\rpbeb}{r_{\bar{p}\bar{e}}}
\newcommand{\rpeb}{r_{p\bar{e}}}
\newcommand{\rpbe}{r_{\bar{p}e}}
\newcommand{\reeb}{r_{e\bar{e}}}

\begin{abstract}
   Adopting explicitly correlated \KW-type basis functions, the Born-Oppenheimer 
   potential curves of a number of excited $\Sigma$ states of the 
   hydrogen-antihydrogen system (\HHbar) were calculated for both, even and odd, 
   Q symmetries, including also free positronium states. 
   It is demonstrated that the excited leptonic states support ro-vibrational 
   states with energies close to the ground-state dissociation threshold.
   As a consequence, the excited leptonic states need to be considered 
   in theoretical treatments of ground-state H--\Hbar\ collisions.
\end{abstract}

\maketitle

\section{Introduction}
Antimatter is not only fascinating by itself, but the evident asymmetry of 
matter and antimatter in our universe (together with the absence of a corresponding 
large amount of cosmic background radiation that would be expected from the 
annihilation of matter and antimatter when generated in equal amounts at the 
big bang) motivates experiments generating and trapping the simplest neutral 
compound antimatter atom, antihydrogen (\Hbar) 
\cite{anti:baur96,anti:blan98,anti:amor02,anti:gabr02,anti:gabr02a}. 
These experiments investigate possible violations of the charge-parity-time (CPT) 
conservation or the weak-equivalence principle (WEP) by comparing the spectroscopic 
properties or gravitational behavior of \Hbar\ with the ones of H, see, 
e.\,g., \cite{anti:bert15,anti:ahma18,anti:bake25,anti:ande23} and references therein. 

With the experimental accessibility of the \Hbar\ atom, 
there is on the other hand also an increasing interest in its interaction 
with ordinary matter. In fact, long before the first experimental realization 
of antihydrogen atoms, the interaction with its matter counterpart, the hydrogen atom, 
has been studied very carefully theoretically \cite{anti:kolo75}. It was found 
that the Born-Oppenheimer Hamiltonian shows a symmetry closely related to 
charge-parity symmetry, called Q symmetry, which splits the eigenvalue spectrum 
into even and odd states. Using explicitly correlated basis functions in 
prolate spheroidal coordinates, known as \KW\ basis functions, a very accurate 
ground-state potential curve was found, proving the local maximum found in 
\cite{anti:morg73} to be an artifact of the limited accuracy of the therein 
adopted numerical approach. Similar to the corresponding anion (cation) 
that is obtained, if one positron (electron) is removed from the \HHbar\ system, 
it was found that the \HHbar\ molecule is stable only above a critical 
distance $R_c$ (also known as Fermi-Teller radius, as it was 
first mentioned in \cite{anti:ferm47} for the corresponding one-lepton ion).  
Below $R_c$ the system breaks-up into protonium (p$\bar{\mathrm{p}}$) and 
positronium (e$\bar{\mathrm{e}}$), since the energy of p$\bar{\mathrm{p}}$ 
and e$\bar{\mathrm{e}}$ lies below the one of \HHbar, at least within 
the Born-Oppenheimer approximation. The occurrence of a critical 
distance is intuitively understandable, as at some point the dipole formed by 
the proton and the antiproton is insufficient to bind the positronium (in the 
united-atom limit the positronium is evidently completely unbound). (A more 
stringent mathematical argument is given in \cite{anti:grid05}.) However, 
the rather sharp change of the slope of the leptonic energy curve close to 
the critical distance lead to some speculation. It may be caused by an avoided 
crossing between the molecular state and the one describing a free 
protonium-positronium pair, or the molecular state simply changes continuously 
into an unbound state \cite{anti:saen06,anti:jons08}.

Besides the importance of \HHbar\ as a paradigmatic matter-antimatter system, 
the interest in this system was also raised by a proposal to use ultracold H 
atoms for the sympathetic cooling of \Hbar\,\cite{anti:shly93} and the 
corresponding experimental progress in cooling H atoms, even achieving 
a Bose-Einstein condensate \cite{cold:frie98a}. This initiated theoretical 
investigations regarding the question whether the elastic cross-section 
responsible for cooling can successfully compete with the destructive 
inelastic collisions, including the rearrangement into protonium and 
positronium and, of course, leptonic or hadronic annihilation 
\cite{anti:voro98,anti:froe00,anti:sinh00,anti:jons01,anti:armo02,
anti:froe04,anti:jons04,anti:zyge04,anti:armo05,anti:jons05,anti:cohe06,
anti:jons06} 
(see \cite{anti:jons18} for a short review), but also, e.\,g., a 
study investigating radiative association \cite{anti:zyge01}.

Most of those scattering calculations were based on the Born-Oppenheimer 
potential curve (possibly including adiabatic corrections) describing the 
interaction of H and \Hbar, each being in their respective ground states. 
This raised the question how to treat the system for interhadronic separations 
$R<R_c$. In fact, adopting explicitly correlated Gaussians Strasburger did 
not only calculate the to the authors' knowledge still most accurate 
ground-state potential curve \cite{anti:stra02}, but later demonstrated a 
complete break-down of the Born-Oppenheimer, even the adiabatic, approximation, 
since the correction terms were shown to diverge \cite{anti:stra04} as the 
interhadronic distance $R$ approaches the critical value $R_c$ (from above).  
Most problematic is thus the treatment of the rearrangement channel into 
protonium and positronium with the question of a proper theoretical framework 
for $R<R_c$. In fact, since the protonium (and in principle also the positronium) 
can be left in different bound states, various rearrangement channels 
are accessible. Furthermore, the two particles (protonium and positronium) 
may have variable translational energy. The results obtained within the 
plane-wave \cite{anti:sinh02} or distorted-wave \cite{anti:jons04} 
approximations were thus found to be improved, if the rearrangement 
channels were explicitly included using the Kohn variational 
method \cite{anti:armo02} or an optical potential \cite{anti:zyge04}. 
In yet another approach, a mass-scaled Born-Oppenheimer potential 
was used to cope with the divergence close to the critical distance 
within a Born-Oppenheimer framework \cite{anti:steg12}. 
In view of the obtained results, in \cite{anti:jons08} the importance 
of the explicit inclusion of the rearrangement channels, e.\,g., in 
a close-coupling type approach (see Eq.\,(7) in \cite{anti:jons08}) 
was emphasized. A general formulation of the required $T$ matrix 
elements including the approximation to include only a single 
molecular Born-Oppenheimer potential is given in \cite{anti:armo07}. 
In \cite{anti:voro08} an approximation was formulated and adopted 
that requires the channels only in the asymptotic regime, 
i.\,e.\ for $R>R_c$, by invoking unitarity arguments. A natural, 
but computationally very demanding extension is to directly treat 
the full four-body problem, as was done in 
\cite{anti:pisz14a,anti:pisz14b,anti:steg16,anti:yama18b}. 
Even if a very powerful four-body code, in this case one based on 
the Gaussian expansion method using different sets of Jacobian 
coordinates, exists \cite{anti:hiya03}, the calculation is extremely 
challenging, since the collision channel of hydrogen and antihydrogen 
involves very highly excited states, even if both atoms start in their 
ground state. The reason is that within a full four-body treatment, 
all the above-mentioned protonium and positronium channels are 
automatically included, starting with their ground states. The  
protonium ground-state energy is (within the non-relativistic 
approximation) $E_{p\bar{p}} = -m_p / 4 \approx \au{-459.038}$, but 
the dissociation energy of \HHbar\ is within the same approximation 
\au{-1.0}. (Atomic units Hartree are used throughout this work.) 
Using the complex-scaling method, scattering resonances close to 
the dissociation threshold were found that in an adiabatic picture 
would be associated to the protonium states with principal quantum 
numbers as high as 22 and 24 \cite{anti:steg16,anti:yama18b}.

Despite the mentioned numerous studies of collisions between 
H and \Hbar\ or studies of the \HHbar\ quasimolecule 
in \cite{anti:labz05,anti:ferr08}, to the authors' knowledge 
all adiabatic calculations 
considered only the leptonic ground-state of \HHbar\ and its 
coupling to the protonium-positronium rearrangement channel. In fact, 
only in \cite{anti:shar06,anti:shar06a} Born-Oppenheimer potential curves
for excited states appear to have been reported at all, but only 
for 4 non-$\Sigma$ (Q-even) states. On the other hand, the present work 
investigates the excited \HHbar\ states of $\Sigma$ symmetry, 
considering both, even and odd Q symmetries. Because, unlike for 
ordinary matter system like $\mathrm{H}_2$, the interaction between 
antiproton and proton is attractive, for small inter-hadronic 
separations this Coulombic attraction dominates the Born-Oppenheimer 
potentials of all leptonic states. Consequently, even in the limit of 
zero collision energy those excited states are energetically 
accessible to an antihydrogen atom scattering on a hydrogen atom, 
both being in their ground states. This allows for resonances 
to occur in which the incident atoms become energetically degenerate 
with some leptonically excited rovibrational molecular bound state. 
It is shown that despite the uncertainty of the proper treatment 
of the system below the critical distance within the here adopted 
Born-Oppenheimer approximation, there is a plethora of rovibrational 
states that can lead to scattering resonances close to the ground-state 
dissociation limit of approximately \au{-1.0}. Noteworthy, even if 
the kinetic energy of the colliding atoms would not perfectly agree 
with the excited molecular bound states, the resonance condition 
maybe fulfilled, since also those bound states have a finite life time 
and thus a non-vanishing energy width due to annihilation and decay 
into protonium and positronium. Therefore, refined scattering 
calculations including the excited \HHbar\ states need to be performed, 
if reliable cross-sections should be obtained. This conclusion is 
further supported by the findings of the full four-body calculations 
in \cite{anti:steg16,anti:yama18b} in which a resonance with dominant 
contribution from an excited positronium state ($n=2$) appears to be 
relevant. As an interesting side aspect it is demonstrated in the 
present work that (discretized) free-positronium states can be obtained 
also when using \KW-type basis functions, if a basis set with two sets 
of non-integer exponents is used. This opens the way to corresponding 
close-coupling scattering calculations in which the rearrangement 
channels are intrinsically included. This avoids the use of two types 
of basis functions, one describing the molecular leptonic states and 
one the (free) positronium that are computationally very demanding.

\section{\HHbar\ and Q symmetry}
The \HHbar\ system consists of a hydrogen and anti-hydrogen atom and is therefore
the simplest neutral matter-antimatter molecule that can be studied. Its leptonic 
Hamiltonian is given by
\begin{equation}
      \hat{H} = -\frac{1}{2}\nabla_e^2 - \frac{1}{2}\nabla_{\bar{e}}^2 
      - \frac{1}{\rpe} - \frac{1}{\rpbeb} - \frac{1}{\reeb} 
      + \frac{1}{\rpeb} + \frac{1}{\rpbe} \text{,} \label{eq:leptonic_hamiltonian}
   \end{equation}
where $r_{ab}$ denotes the distance between constituents and
$p$ and $e$ denote the proton and electron, respectively, while 
$\bar{p}$ and $\bar{e}$ denote their antimatter counterparts. 
Evidently, the Hamiltonian of \HHbar\ differs from its 
corresponding matter counterpart H$_2$ only by an exchange of signs.

In \cite{anti:kolo75} Ko{\l}os et al.\ introduced the $\hat{Q}$ symmetry operator,
a composition of mirroring the leptons on a plane bisecting the inter-hadronic
axis ($\hat{\sigma}_{\rm lep}$) and permuting the leptons ($\hat{P}_{e\bar{e}}$).  
Both operations alone do not commute with the leptonic Hamiltonian, 
\begin{equation}
      \hat{\sigma}_{\rm lep}\hat{H} = -\frac{1}{2}\nabla_e^2 - \frac{1}{2}\nabla_{\bar{e}}^2 - \frac{1}{\rpbe} - \frac{1}{\rpeb} - \frac{1}{\reeb}+ \frac{1}{\rpbeb} + \frac{1}{\rpe} 
   \end{equation}
and   
   \begin{equation}
      \hat{P}_{e\bar{e}}\hat{H} = -\frac{1}{2}\nabla_e^2 - \frac{1}{2}\nabla_{\bar{e}}^2 + \frac{1}{\rpeb} + \frac{1}{\rpbe} - \frac{1}{\reeb} - \frac{1}{\rpe} - \frac{1}{\rpbeb} ,
   \end{equation}
but leave the kinetic and $1/{\reeb}$ terms unchanged while flipping the
sign of all remaining terms. The Hamiltonian therefore is unchanged under 
the Q symmetry operation. This symmetry is not observed in ordinary matter 
molecules and closely related to charge-parity (CP) symmetry. However, 
the leptonic Hamiltonian of \HHbar\ possesses the same cylindrical symmetry 
around the inter-hadronic axis as any matter diatomic molecule. Therefore, 
the total angular momentum of the electrons $\Lambda$ is a good quantum 
number and one can solve for the $\Sigma, \Pi, \Delta, \dots$ states 
separately. On the other hand, as all 4 particles are different, there 
is evidently no exchange symmetry and thus no ortho and para \HHbar, in 
contrast to the case of H$_2$. 

As was pointed out in \cite{anti:kolo75}, an advantage of the Q symmetry 
is that it is very suitable for calculations within the Born-Oppenheimer 
approximation, since within this approximation the Hilbert space of \HHbar\ can 
be separated into Q-even and Q-odd leptonic wave functions. Thus, 
symmetry adapted basis functions may be used to reduce the computational 
efforts. Furthermore, as long as the adiabatic approximation is a good 
approximation, selection rules with respect to leptonic transitions 
may be formulated. These may, of course, be completely different from 
the ones known for ordinary matter. For example, even the ground state 
with $\Sigma$ symmetry (Q even) possesses a non-vanishing dipole moment 
allowing, e.\,g., for radiative transitions in between rovibrational 
states belonging to the same leptonic state \cite{anti:zyge01}.

\section{Numerical Approach}
Despite being the simplest neutral matter-antimatter molecule, \HHbar\ is
still a four-body problem. Exact solutions are therefore difficult to obtain,
especially for (highly) excited states. Therefore, the Born-Oppenheimer 
approximation is employed in which the wave function
\begin{equation}\label{eq:Psi}
   \Psi_{m,\nu,J}({\bf r}_p, {\bf r}_{\bar{p}}, {\bf r}_e, {\bf r}_{\bar{e}}) 
        = \chi_{\nu,J,m_J}^{(m)}({\bf R}) \; \psi_m({\bf r}_{e}, {\bf r}_{\bar{e}}; R)  
\end{equation}
is assumed to be a single product wave function of the hadronic (rovibrational) 
wave function $\chi_{\nu,J,m_J}^{(m)}$ that depends solely on the inter-hadronic 
separation vector ${\bf R}={\bf r}_p - {\bf r}_{\bar{p}}$ and the leptonic wave 
function $\psi_m$ that depends on the leptonic coordinates, but only 
parametrically on $R=|{\bf R}|$.

\subsection{Leptonic Problem}
The leptonic wavefunctions and corresponding Born-Oppenheimer potential curves 
are obtained with the help of the computer program H2SOLV published by
K.\ Pachucki et al.\ \cite{dia:pach16}. The code, written for diatomic 
two-electron molecules was correspondingly modified for treating \HHbar, 
similar to the modification of the \KW\ code in \cite{anti:froe00}. In more 
detail, for the here considered $\Sigma$ states the leptonic wave function 
$\psi_m$ is expressed as a superposition 
\begin{equation}\label{eq:psi_lep}
   \psi_m ({\bf r}_{e}, {\bf r}_{\bar{e}}; R) 
        = \sum_{\mathbf{n}} c_\mathbf{n} (R) \;
            \tilde{\phi}_\mathbf{n} ({\bf r}_{e}, {\bf r}_{\bar{e}}; R) 
\end{equation}
of so-called \KW\ basis functions
\begin{eqnarray}
   \tilde{\phi}_\mathbf{n} & = &\exp\left (R(-y\eta_1 - x\eta_2 - u\xi_1 - w\xi_2)\right) \times \phantom{xxxxx}\nonumber\\
         & & \phantom{xxx} r_{e\bar{e}}^{n_0}\,\eta_1^{n_1}\,\eta_2^{n_2}\,\xi_1^{n_3}\,\xi_2^{n_4}\,r^{1 - n_0 - n_1 - n_2 - n_3 - n_4}  \label{eq:kw}
\end{eqnarray}
where $\xi_j=(r_{p,j}+r_{\bar{p},j})/R$ and $\eta_j=(r_{p,j}-r_{\bar{p},j})/R$ 
are the prolate-spheroidal coordinates of lepton $j$ 
($\varphi_j$ being its angle around the inter-hadronic axis).  
The index vector $\mathbf{n}$ contains the set of five integers (quintuple) 
$\{n_0,n_1,n_2,n_3,n_4\}$ that defines an individual \KW\ {\em basis function}. 
A {\em basis set} is then characterized by the set of exponential parameters 
$y, x, u$, and $w$ that is identical for all basis functions, being  
henceforth called the {\em base} %
\footnote{The definition of the base parameters of Pachucki et al.\ differs  
slightly from the original one introduced by \KaW\ who used the symbols 
$\{\alpha,\bar{\alpha},\beta,\bar{\beta}\}$. They are related by 
$uR = \alpha$, $wR = \bar{\alpha}$, $yR = -\beta$, $xR=-\bar{\beta}$. 
The motivation for dividing the \KW\ parameters by $R$ is that in this case 
the exponential damping becomes independent of the inter-hadronic separation, 
since the implicit dependence contained in the prolate-spheroidal coordinates 
$\xi$ and $\eta$ is canceled.}, %
and a selection of a set of quintuples $\mathbf{n}$. 

Due to the very limited computational 
resources, \KaW\ selected in their original works on H$_2$ (and later 
for \HHbar) the quintuples ${\bf n}$ by hand. This selection, though 
extremely successful in obtaining very accurate results with astonishingly 
small basis sets, was hampered by linear dependencies between the 
basis functions due to the finite numerical precision. In H2SOLV this 
problem is overcome by adopting FORTRAN libraries allowing for variable 
precision \cite{gen:MPFUN2015}. In this way, systematic convergence 
investigations for a given 
base can be performed by a systematic increase of the number of basis 
functions. In H2SOLV the integer parameter $\Omega$ is introduced 
for a complete characterization of a basis set, as all quintuples ${\bf n}$ 
that fulfill 
\begin{equation}\label{eq:omega}
   \sum_{i=0}^4 n_i \le \Omega
\end{equation}
form a given basis set. A comparison of the results obtained for 
increasing values of $\Omega$ and calculating the relative difference  
to the result obtained with the largest value of $\Omega$ allows for
a systematic convergence test. (Further refinement is possible by 
introducing upper limits for subsets of the $n_j$, e.\,g., separate 
upper limits for $n_0$, $n_1+n_2$, and $n_3+n_4$.)   

Insertion of the {\it ansatz} (\ref{eq:psi_lep})  
for $\psi$ into the leptonic Schr\"odinger equation leads to a generalized 
eigenvalue problem, since the basis functions are non-orthogonal. The 
solution of this eigenvalue problem yields the eigenvector  
$c_m (R)$ and the leptonic energies $E_{{\rm lep}, m}(R)$. 
As in the case of ordinary molecules, the addition of the 
inter-hadronic attraction term, a simple constant, to the 
leptonic energy $E_{{\rm lep}, m}(R)$ yields the Born-Oppenheimer 
potential curve $V_m (R)$ for the leptonic state $m$. 

Since the leptonic Hamiltonian in \eqref{eq:leptonic_hamiltonian}
obeys the Q symmetry as discussed above, the leptonic energy spectrum 
splits into states that are either even or odd with respect to 
Q symmetry. The introduction of the symmetry-adapted linear 
combination of \KW\ basis functions
\begin{eqnarray}
   \phi_\mathbf{n}^{\pm} = \tilde{\phi}_\mathbf{n} 
         \pm (-1)^{n_1 + n_2}\tilde{\phi}_\mathbf{n} \; , \label{eq:salc}
\end{eqnarray}
which is Q even or Q odd, respectively, leads to a substantial reduction 
of the computational time and memory needs. Especially, if many or even 
all eigenstates for a given basis set are requested, this effect is 
highly relevant, since the number of matrix elements that need to be 
calculated and the size of the matrix that needs to be diagonalized is 
reduced by about a factor 2. Remind, that variable precision 
needs to be adopted in order to cope with numerically caused linear
dependencies and this problem becomes more severe for larger basis 
sets (increasing values of $\Omega$). Besides the linear dependencies 
that are due to finite numerical precision, additional true linear 
dependencies occur, if all symmetry adapted basis functions obtained 
according to Eqs.~\eqref{eq:omega} and \eqref{eq:salc} are included 
in the calculation, as this may lead even to zero-valued wave functions 
due to possible symmetries in the base parameters $(y, x, u, w)$. 
These wavefunctions must thus be identified and filtered out in advance, 
as is described in detail in the appendix \ref{app:filter_rules}.

\begin{figure}
   \includegraphics[scale=0.5]{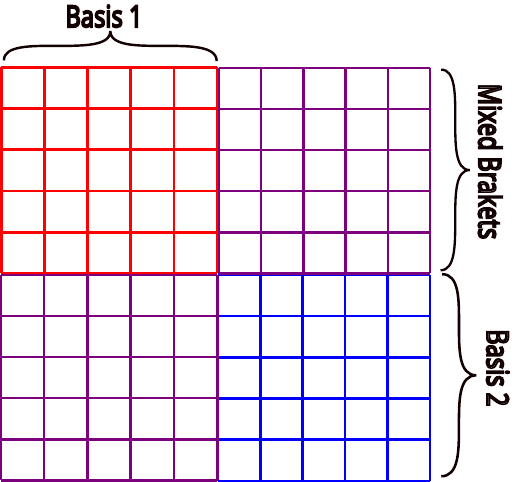}
   \caption{
      Structure of the Hamiltonian matrix for a dual-base basis set, i.\,e.\ a basis 
      set comprising of two basis sets $\{\phi^{(0)}\}$ and $\{\phi^{(1)}\}$ with 
      the different sets of exponential parameters $(y_0, x_0, u_0, w_0)$ and 
      $(y_1, x_1, u_1, w_1)$, respectively. This leads to three types of matrix elements:    
      $\langle\phi_i^{(0)}|\hat{H}|\phi_j^{(0)}\rangle = \langle\phi_j^{(0)}|\hat{H}|\phi_i^{(0)}\rangle$ (red), 
      $\langle\phi_i^{(1)}|\hat{H}|\phi_j^{(1)}\rangle = \langle\phi_j^{(1)}|\hat{H}|\phi_i^{(1)}\rangle$ (blue), and 
      $\langle\phi_i^{(0)}|\hat{H}|\phi_j^{(1)}\rangle \neq  \langle\phi_i^{(1)}|\hat{H}|\phi_j^{(0)}\rangle$ (purple).
   }
   \label{fig:mixed_basis}
\end{figure}

While the \KW\ basis functions in the limit $\Omega\rightarrow \infty$ 
form a complete basis (at least, if the practical problems of requiring 
possibly an infinite precision for reaching this hypothetical limit 
are ignored), the necessary sub Hilbert space required for accurately 
describing a single or a number of states could be covered with much 
less basis functions, if different bases would be adopted. However, 
the historical success of the \KW\ basis functions is due to the 
fact that using a single base allows for the very efficient evaluation 
of the Hamiltonian (and overlap) matrix elements, as they can be obtained 
via recursion from a master integral (${\bf n}=\{0,0,0,0,0\}$). A natural 
compromise in between a single base and, in the extreme case, one 
base per basis function is given by adopting a dual-, triple-, etc. 
base. In this way, also a possible continuous change of the character of 
a leptonic molecular state as a function of inter-nuclear separation can 
be described efficiently. An example would be the admixture of an ionic 
component to a covalent state. Therefore, the here made extension to 
H2SOLV also includes the option to use multiple-base basis 
sets. Such an extension is rather straightforward, as is illustrated in 
Fig.~\ref{fig:mixed_basis} for the case of a dual base that is used 
in the present work. Since the product of two \KW\ basis functions 
is again a \KW\ basis function, only with different base parameters and 
quintuples, a dual base basically increases the number of master integrals 
by factor 3. Of course, if the same $\Omega$ is used as for a single-base 
basis set, the dimension of the Hamiltonian matrix that needs to be 
diagonalized also increases by factor 2. 

\subsection{Optimization of the Base Parameters}\label{sec:optimization}
Highly accurate calculations of leptonic states already of two-lepton 
systems like H$_2$ or \HHbar\ adopting finite basis sets require a 
careful individual optimization of all basis-set parameters for every 
single state. In the case of the lowest state of a given symmetry, 
the variational principle allows for a clear identification of the 
best basis set. Projecting out all energetically lower-lying states, 
the variational principle is also strictly applicable to the excited 
states. If a set of states is obtained by diagonalization of the 
Hamiltonian matrix resulting from the linear 
{\it ansatz} \eqref{eq:psi_lep}, the enforced orthogonality is 
equivalent to projecting out the other states. Thus the variational 
principle would be applicable to all states, but only if the 
representation of all lower-lying states would be exact. Although this 
is evidently not the case, if a finite basis set is adopted like the 
one in this work, the variational principle is usually still reasonably 
well fulfilled, at least for the lower-lying excited states. Since the 
aim of this work is a reasonably good description of the lower lying 
leptonic states, but not an extremely highly accurate description 
of individual states, a dual base with $\Omega$ values up to 10 was 
found to be a good compromise between required computational resources 
and achieved accuracy.

The choice of at least two bases, i.\,e.\ a dual-base basis set was also 
motivated by the fact that from the previous results it is evident that 
the leptonic states of \HHbar\ should comprise at least one molecular 
and one positronium component. Furthermore, the relative contribution of 
them should change with varying inter-hadronic separation. This 
information could have been used when optimizing the base parameters of 
the two bases. For example, one base could have been optimized by 
minimizing the \HHbar\ ground-state energy, while the other base minimizes  
the energy of a free, more accurately "box" discretized  positronium that 
would be obtained by setting both hadronic charges to 
zero \cite{anti:froe00,anti:jons01}. However, since the present work 
aims primarily at a (reasonably) good description of a larger number 
of Born-Oppenheimer curves, a different strategy was pursued that, after 
numerous trials, was found to be more suitable for the present purpose. 

The idea was that one base should provide a good description for large  
inter-hadronic distances where the atomic or molecular character should 
dominate, and one for small distances where the positronium character 
is expected to contribute, too. The most straightforward way would be 
to try to optimize all 8 base parameters simultaneously, using as a criterion 
the energy minimization of a large number of states at different 
internuclear separations. However, this would not only pose a very 
challenging, if not with the available computational resources 
impossible task, but would also require to define a good minimization 
criterion. Evidently, for states with very different character it can 
easily occur that their energy ordering switches, depending on the 
base parameters. This effect is enhanced by the fact that with increasing 
energy the density of states also increases drastically. This leads to 
a very complicated, fine-structured optimization landscape with many 
local minima. 

In order to avoid such complications, the parameters of the two bases 
(separately determined for both symmetries Q even and odd) were found  
independently, already reducing the number of parameters in each 
optimization from 8 to 4, leading to a dramatic reduction of the 
computational efforts. Furthermore, one base was solely optimized 
by minimizing the energy of the lowest lying state (in the case of 
Q-even symmetry the ground state) at a single inter-hadronic separation 
($R=5.0\,a_0$). A four-dimensional gradient descent algorithm was used 
to find the corresponding base parameters $\{y_0, x_0, u_0, w_0\}$. 
Since at large inter-hadronic separations as $R=5.0\,a_0$ the 
ground state is known to behave, in a very good approximation, 
almost as two indpendent atoms \cite{anti:saen06} that are described 
by the base parameters $\{0.5, -0.5, 0.5, 0.5\}$, these values were used 
as starting values for the optimization. The lowest ground-state 
energy at $R=5.0\,a_0$ found this way with $\Omega=10$ was obtained 
with the base parameters listed in the first row of 
Table \ref{tab:optimised_basis_parameters}. The choice of the 
start parameters of the second base that should provide a good 
description for small inter-hadronic separations is less straightforward, 
if one does not want to represent a free prositronium, but the excited 
molecular states. After some trials, the same start parameters as for 
the first base (representing indpendent H and \Hbar\ atoms) were also 
used as starting point for the second base $\{y_1, x_1, u_1, w_1\}$. 
However, now the energy of the 2nd excited state is optimized. 
Most importantly, the optimization starts at $R=5.0\,a_0$, but 
the optimal parameters found at this value of $R$ only serve as 
a starting point to find the optimal parameters at $R=0.8\,a_0$. 
The latter are found  by decreasing $R$ in small steps and using 
the corresponding optimized parameters as starting point (for the 
four-dimensional gradient descent algorithm) at the next $R$ value. 
With the aid of this iterative procedure, finally the optimized 
base parameters $\{y_1, x_1, u_1, w_1\}$ at $R=0.8\,a_0$ listed 
in the 2nd row of Table \ref{tab:optimised_basis_parameters} 
were determined. An analogous procedure was applied in order 
to find the optimized parameters of the two bases used in the 
calculations of the states with Q-odd symmetry, listed in the 
3rd and 4th rows of Table \ref{tab:optimised_basis_parameters}. 
However, in this case the 1st and not the 2nd excited state 
was used when determining the base parameters of the 2nd basis. 
It turns out that the introduction of the 2nd base that is optimized 
for the excited state is crucial for obtaining also free positronium 
states as is discussed in more detail below.

\begin{table}
   \caption{Optimized base parameters found by the procedure discussed in 
            Sec.\,\ref{sec:optimization} and used in the calculations 
            of this work.}
   \center
   \begin{tabular}{c|c|c|c|c|c|c}
      R & Symmetry & State & y & x & u & w\\
      \hline
      5.0 & Q-even & 0 & 0.3 & -0.5 & 0.5 & 0.5 \\
      \hline
      0.8 & Q-even & 2 & -0.625 & -0.65 & 0.121875 & 0.14375 \\
      \hline
      5.0 & Q-odd & 0 & -0.1 & -0.5 & 0.3 & 0.5 \\
      \hline
      0.8 & Q-odd & 1 & -3.86875 & -3.2825 & 0.14375 & 0.1375
   \end{tabular}
   \label{tab:optimised_basis_parameters}
\end{table}

\subsection{Hadronic Problem}
Once the leptonic problem is solved the resulting Born-Oppenheimer 
curves can be used to solve the hadronic Schr\"odinger equation,  
yielding the total rovibronic (rotational, vibrational, and 
leptonic) energies of the \HHbar\ molecule within the 
Born-Oppenheimer approximation. 

In the absence of an external potential the Hamiltonian for the 
nuclear motion is isotropic and thus the angular solutions of 
$\chi_{\nu,J,m_J}^{(m)}(\vec{R})$ in Eq.\,\eqref{eq:Psi} are 
given by the spherical harmonics $Y_{J,m_J}(\theta_R,\varphi_R)$ 
with the two angles $\theta_R$ and $\varphi_R$ describing the 
orientation of the inter-hadronic separation vector ${\bf R}$. 
$J$ is the rotational quantum number, every rotational state 
being $2J+1$ times degenerate. The remaining one-dimensional 
radial equation describing vibrational (or dissociative) motion 
within one Born-Oppenheimer potential $V_m(R)$ can be solved 
numerically in various ways. In this work the radial wavefunction 
is expanded in a $B$-spline basis yielding a generalized eigenvalue 
problem. An advantage of the basis-set expansion is that, depending 
on the adopted number of $B$ splines a large number of rovibrational 
solutions are obtained simultaneously. 2048 $B$ splines of order $k=15$, 
distributed over a linear knot sequence in between $R_0=0\,a_0$ and 
$R_{\rm max} = 5.0\,a_0$, were found to provide sufficiently converged 
results. 

Within the here adopted Born-Oppenheimer approximation there is 
the intricate question how the potentials curves should be continued 
below the critical distance, since the non-adiabatic corrections 
diverge \cite{anti:stra04}, as was already mentioned. This is 
evidently directly connected to the question, whether the critical 
distance is caused by a Feshbach-type resonance between a molecular 
state and the free positronium-protonium pair or by a continuous 
change of the state \cite{anti:stra04,anti:saen06,anti:jons08}. Adopting 
the latter assumption, every Born-Oppenheimer potential represents one 
state completely and its leptonic energy $E_{{\rm lep},m}(R)$ should 
converge for small values of $R$ to the energies of a free positronium. 
The leptonic energies were thus correspondingly extrapolated below 
$R=0.8\,a_0$. This value slightly above the most accurate value reported 
for the critical distance ($R_c=0.744\,a_0$ \cite{anti:stra02}) was 
chosen for two reasons. First, the in this work adopted basis sets 
start to deviate visibly from the best available ground-state curve 
in \cite{anti:stra02} at around this value of $R$. Second, it was 
concluded in \cite{anti:stra04} that the adiabatic approximation should 
become inadequate in between $R=0.8\,a_0$ and $1.0\,a_0$. The 
sensitivity of the obtained rovibronic energies on 
the exact form of the extrapolation was checked by testing rather 
different functional forms and was found to be very small. 
In view of the adopted approximations, starting with the 
Born-Oppenheimer approximation, the uncertainty induced by the 
extrapolation below the critical value appears thus to be negligible, 
since the present work does not strive for spectroscopic data for 
single rovibronic states, but investigates the density of rovibronic 
states close to the dissociation threshold of the leptonic ground 
state. Thus, in the end, a simple linear extrapolation was used 
for the results shown in this work, but in Fig.~\ref{fig:rovib_states_scatter} 
a comparison of the results obtained with either linear or a 6th order 
polynomial are also given.

An additional practical problem arises, because the Born-Oppenheimer 
potentials are precalculated at a finite number $N$ of inter-hadronic separations 
$R_i, i=1,N$. However, the numerical integration that is performed when solving the 
Schr\"odinger equation, in the present case when setting up the Hamiltonian 
matrix in the $B$-spline basis, typically requires the values at arbitrary 
$R$ values (within the the interval $[R_0,R_{\rm max}]$). The 
interpolation in between the the discrete set of $R_i$ values is 
usually rather straightforward and in the present approach 
a cubic spline interpolation is adopted. However, from the attractive 
inter-hadronic interaction term ($V_{p,\bar{p}}=-1/R$) in the leptonic 
Hamiltonian it is clear that all \HHbar\ potential curves $V_m(R)$ 
diverge for $R \rightarrow 0\,a_0$. This evidently leads to numerical problems, 
if the integrals of the potential-energy operator between the $B$ splines 
close to $R_0=0\,a_0$ are calculated. Stable solutions were found when 
using the function 
\begin{equation}
    V_{\rm ext}(R) = - A - B \, \exp(-C R)
\end{equation}
within the interval $[R_0,10^{-6}\,a_0]$. The same coefficients 
($A=4.545473\times 10^{4}$, $B=1.294251\times 10^{5}$, and $C=3.044522\times 10^{6}$) were 
used, since in that small $R$ range close to zero the function is completely 
dominated by the inter-hadronic interaction term that is identical for 
all states. It was checked that the artifical extrapolation that enforces 
a finite value at $R\rightarrow 0\,a_0$ has a negligible effect on the 
calculated energies that is much smaller than the already small impact 
of the extrapolation in the interval $[10^{-6}\,a_0,0.8\,a_0]$. The 
reason is that all wavefunctions are localized at much larger values 
of $R$ than $R=10^{-6}\,a_0$.

\section{Results}
\subsection{Leptonic energy spectrum}\label{sec:spectrum}
\begin{figure*}
   \includegraphics[scale=1]{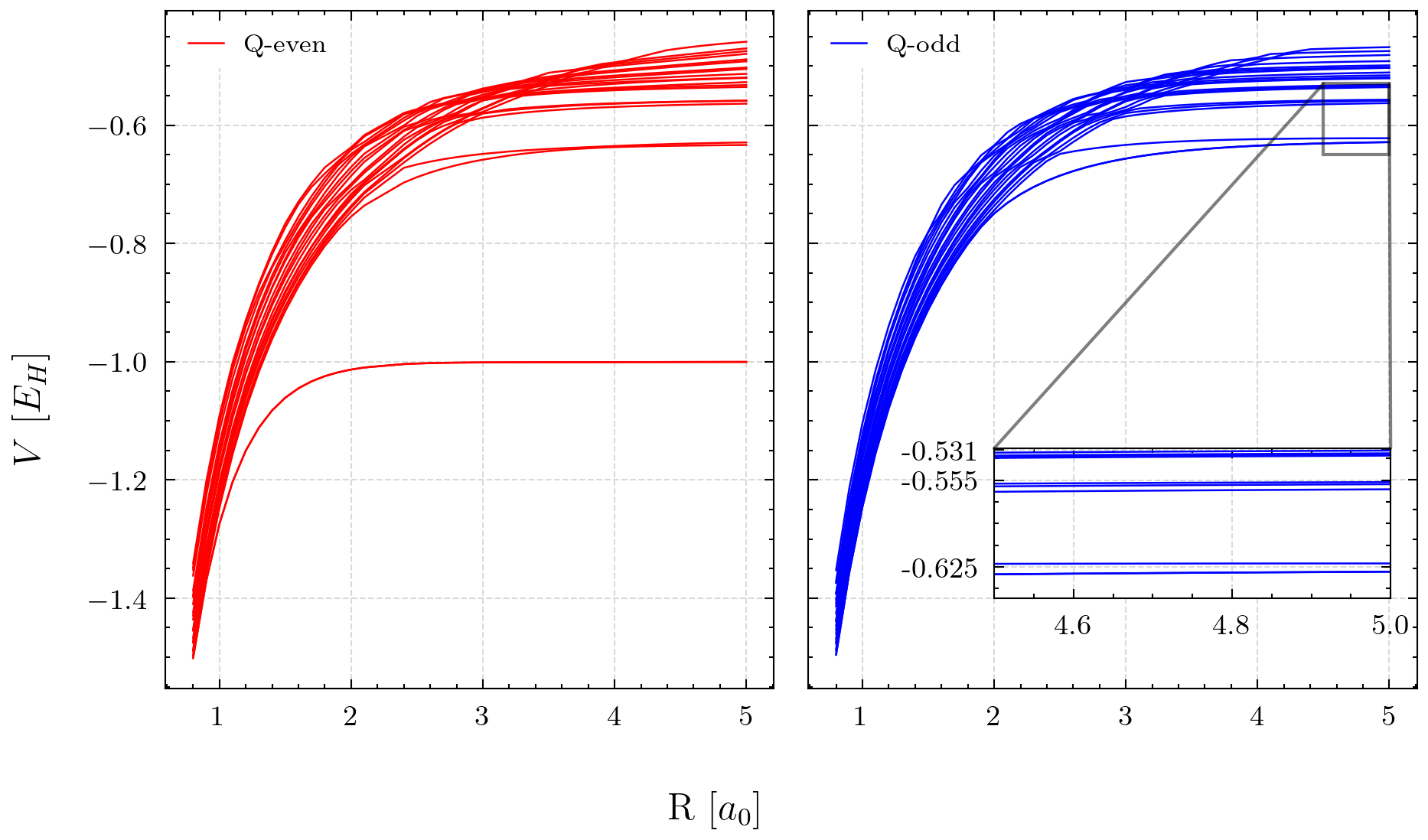}
   \caption{
      Lowest-lying Born-Oppenheimer potential curves of \HHbar\ (in Hartree) as 
      a function of the inter-hadronic separation $R$ (in Bohr) for 
      the $\Sigma$ states that possess Q-even (left, red) or Q-odd (right, 
      blue) symmetry. The inset shows a zoom that more clearly reveals the 
      asymptotic behavior of the low-lying Q-odd states. 
      (The data points are connected by lines for guiding the eye.  
      The basis-sets described in Sec.~\ref{sec:optimization} with 
      $\Omega=10$ were adopted.)
   }
   \label{fig:bo_overview}
\end{figure*}

The Born-Oppenheimer potential curves $V$ obtained using the bases in 
table \ref{tab:optimised_basis_parameters} and $\Omega=10$ are 
presented in Fig.~\ref{fig:bo_overview} for both Q-even and 
Q-odd states, though separately. (The numerical values of the 10 lowest states 
of each symmetry are given in the Tables \ref{tab:bo_qpos} and 
\ref{tab:bo_qneg} of Appendix \ref{sec:BO_curves}.)
Expectedly, all states are clearly dominated by the attractive 
inter-hadronic potential and thus basically go as $-1/R$ 
to $-\infty$ as $R$ approaches zero. This is equally true for 
both, Q-even and Q-odd states. Therefore, all states should 
support (numerous) rovibrational states where the low-lying 
ones should be very similar to protonium states. This finding 
contradicts the claim in \cite{anti:shar06} where it is stated 
that the potential curves corresponding to odd states need not 
to be considered, since they "increase monotonically as 
the internuclear distance $R$ decreases". Therefore, Q-odd states 
cannot {\it per se} be neglected in collision processes, except 
for reasons of symmetry conservation. 

In the limit of separated atoms there are $2n$ degenerate states
each with the energy of $-(1 + 1/(n+1)^2)/2$, which are the energies
of one of the (anti-)atoms being in the ground state and 
one (anti-)atom being excited to a state with principal quantum 
number $n$. The leptonic potential curves approach these energies 
with the expected degeneracies as can be seen in the inset of 
Fig.~\ref{fig:bo_overview}.

\begin{figure*}
   \includegraphics[width=\textwidth]{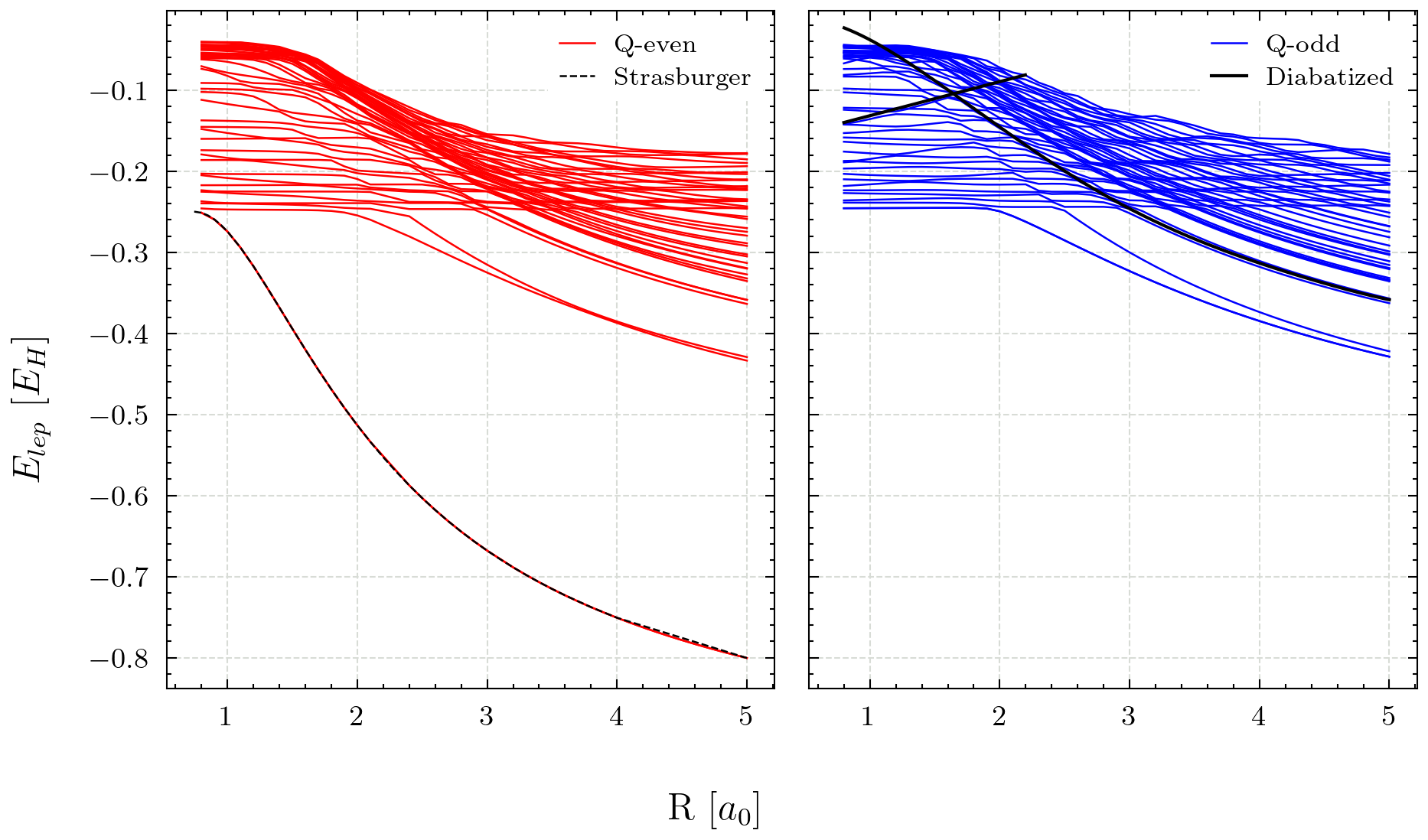}
   \caption{
      As Fig.~\ref{fig:bo_overview}, but showing the leptonic energies 
      $E_{\rm lep}$ instead of the potential curves $V$. Also shown 
      is the ground state obtained by Strasburger \cite{anti:stra02} 
      (left, black dashed) and two diabatized states, sketched  
      manually for guiding the eye (right, black solid).
   }
   \label{fig:leptonic_overview}
\end{figure*}

\begin{figure}
   \includegraphics[width=\columnwidth]{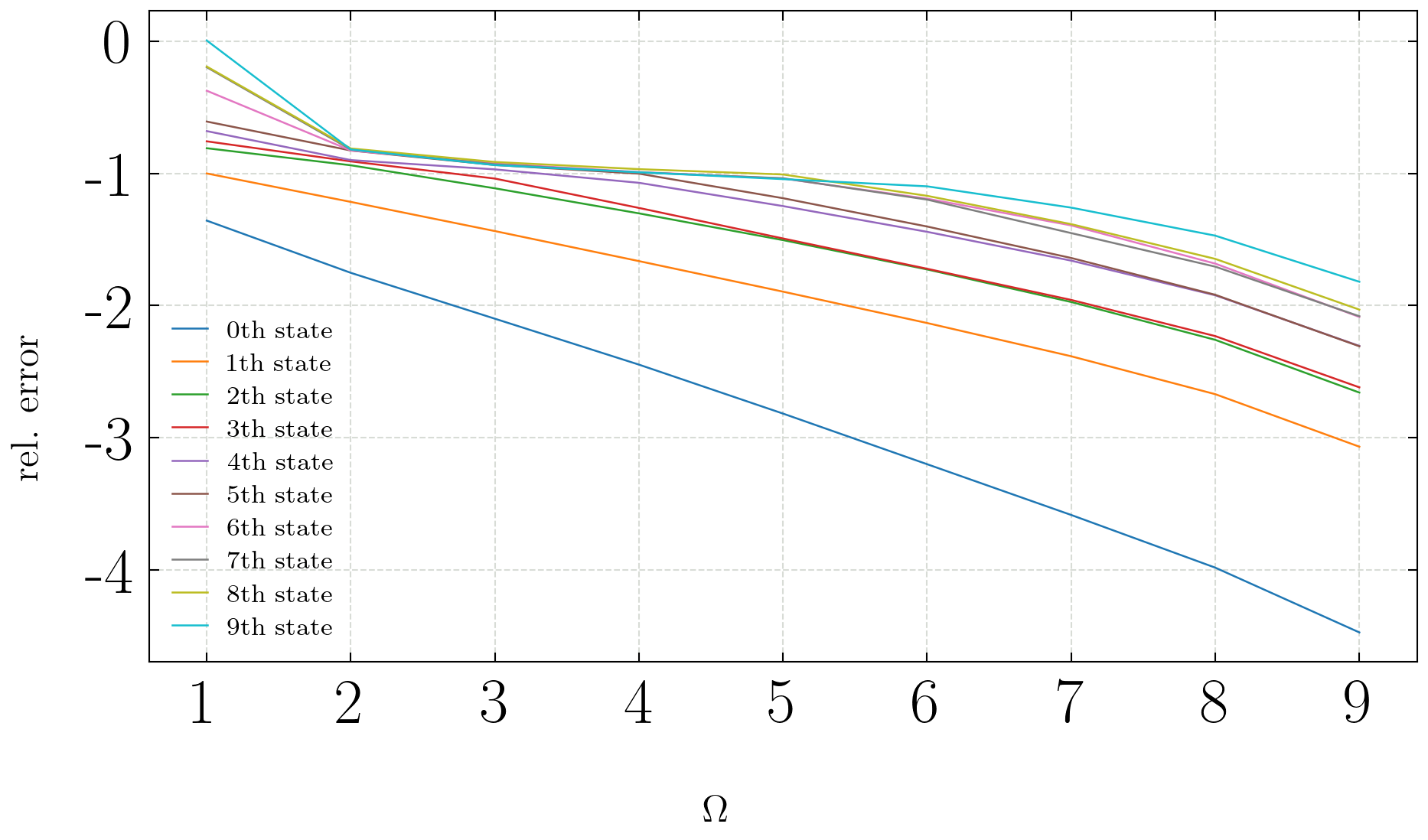}
   \caption{
      Convergence study for the 10 lowest-lying Q-even $\Sigma$ states 
      and $R=0.8\,a_0$. 
      The basis-set is sytematically improved by increasing $\Omega$, 
      see Eq.~\eqref{eq:omega}. Shown is the relative error with respect to 
      the results obtained with $\Omega=10$.}
   \label{fig:conv08}
\end{figure}

\begin{figure}
   \includegraphics[width=\columnwidth]{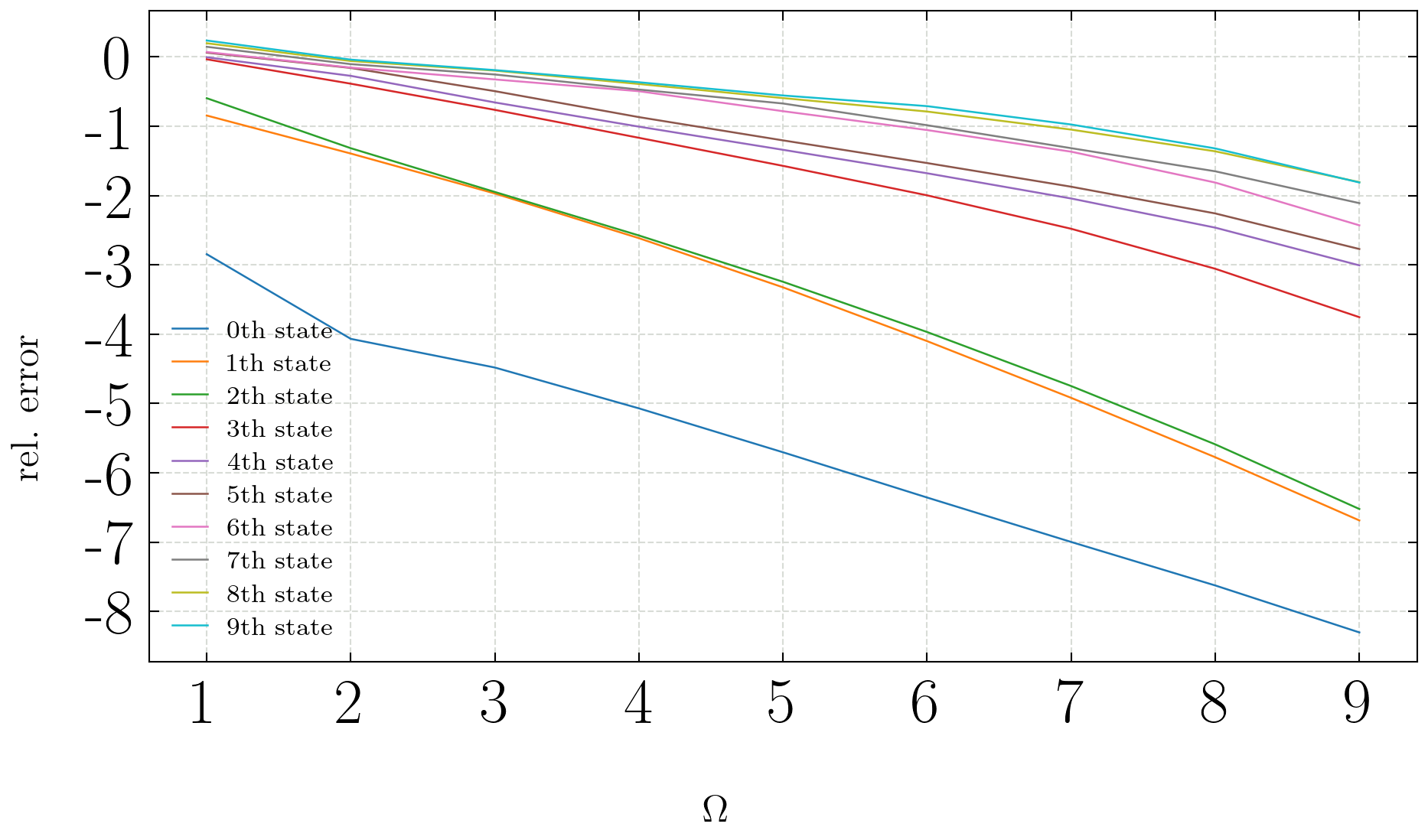}
   \caption{
      As Fig.~\ref{fig:conv08}, but for $R=5.0\,a_0$.}
   \label{fig:conv50}
\end{figure}

A clearer picture is obtained, if the dominant $-1/R$ 
term is subtracted and the leptonic energy curves $E_{\rm lep}$ 
are considered, see Fig.~\ref{fig:leptonic_overview}.
Also shown is the to the authors' knowledge most accurate ground-state 
potential curve \cite{anti:stra02} to which the present result 
agrees very well: the relative deviations are $3.7\times 10^{-5}$ at 
$R=0.8\,a_0$ and $4.3\times 10^{-8}$ at $R=5.0\,a_0$. (Due to the 
more sparse data set in \cite{anti:stra02} and the linear 
interpolation it falsely appears as the present 
results are lower in between $R=4.0$ and $5.0\,a_0$, but they 
lie consistently above.) Since there are no further data 
available to which the excited-state results can be compared, 
convergence studies with respect to $\Omega$ were performed. 
They are shown for $R=0.8\,a_0$ and $5.0\,a_0$ in 
Figs.~\ref{fig:conv08} and \ref{fig:conv50}, respectively. 
\footnote{Convergence plots for additional $R$ values and also for 
the 10 lowest lying Q-odd states are given in Appendix \ref{sec:BO_curves}.}
At least for sufficiently large values of $\Omega$ an exponential 
decrease of the error can be seen. While the ground state is 
converged up to the $\mathcal{O}(10^{-3})$ for inter-hadronic 
distances below $R=2.0$ and up to $\mathcal{O}(10^{-7})$ 
for $R>2.0$, higher excited states are less and less well converged.
This, as well as the later onset of the exponential improvement 
is somehow expected, since with the level of excitation the number 
of nodes increases. Their proper representation requires correspondingly 
polynomials of higher order and thus larger $\Omega$ values.

The leptonic energy spectrum clearly shows a large number of avoided 
crossings between two kinds of states. In the case of Q-even symmetry, 
these avoided crossings are clearly visible starting from the 3rd state 
(the numbering refers to the $R\rightarrow \infty$ limit). In the case 
of Q-odd symmetry, the visibility starts already at the 2nd state. In fact, 
leaving out the ground state, the spectra for Q-even and Q-odd 
states look on the first glance very similar. States that decrease in 
energy for increasing inter-hadronic separation $R$ are repeatedly 
`crossed' by states with almost constant energy leading to the 
large number of avoided crossings. Noteworthy, the slope of the 
$R$-dependent states is also very similar, only slowly decreasing with 
increasing excitation $n$. As for the ground state, the increasing 
energy for decreasing $R$ is a consequence of the fact that 
as H and \Hbar\ approach each other, the energy gain due to the 
larger overlap of the wavefunctions of the electron and the positron 
is overcompensated by the repulsion of the electron (positron) and  
the antiproton (proton), and thus the reduced binding energy to the 
hadrons. Evidently, the almost constant energies observed for the 
other manifold of state as a function of $R$ and thus the independence 
on the $R$ is a strong indication that these states represent (discretized)  
free positronium states. A further validation of this  interpretation is 
given in the following Sec.~\ref{sec:positronium}.

Interestingly, the Q-odd manifold also contains states that, diabatically 
continued, increase in energy, as $R$ increases. This is clearly in   
contrast to the behavior of the other states. As a guide to the eye, 
the diabatic state with the largest slope is highlighted (black line), 
together with a state that shows the "usual" behavior as is known from 
the ground state. A more detailed analysis beyond the scope of this 
work would be needed in order to find a possible explanation (and to 
rule out that they are just an artifact of an insufficient basis set).

\subsection{Discretized Positronium States}\label{sec:positronium}
As was mentioned before, earlier studies have shown that the theoretical 
description of collisions between antihydrogen and hydrogen require the 
inclusion of states that describe the rearrangement into positronium 
and protonium, optimally in a close-coupling framework. In the earlier 
studies of these collisions using \KW\ wavefunctions the calculations 
did not show signatures of states describing free positronium. Therefore, 
$T$-matrix elements describing the rearrangement channel were calculated 
by obtaining the wavefunctions in the final positronium channel setting 
the hadronic charges to zero \cite{anti:froe00,anti:jons01}. Alternatively, 
the positronium wavefunction was obtained numerically \cite{anti:zyge04}. 
Clearly, it would be very attractive, to obtain the wavefunctions of the 
rearrangement process, i.\,e.\ the molecular \HHbar\ states as well as 
the (discretized) free positronium states within a single diagonalization 
of the complete Hamiltonian, as is the case in full four-body calculations. 
As was discussed in Sec.~\ref{sec:spectrum}, the present calculation 
does not only reveal states that show the expected behavior of molecular 
\HHbar\ states (similar to the ground state), but also states with energies 
almost independent of $R$ which is a strong indication that they are, in 
fact, the sought for positronium states. Noteworthy, adopting only a 
single-base basis set, no constant-energy states were observed, as in 
earlier calculations adopting \KW\ basis functions. While these two 
findings ($R$-independent energy and the appearance only with the use 
of a dual-base basis set) is already a strong indication, a further 
validation was made. 

\begin{figure}
   \includegraphics[width=\columnwidth]{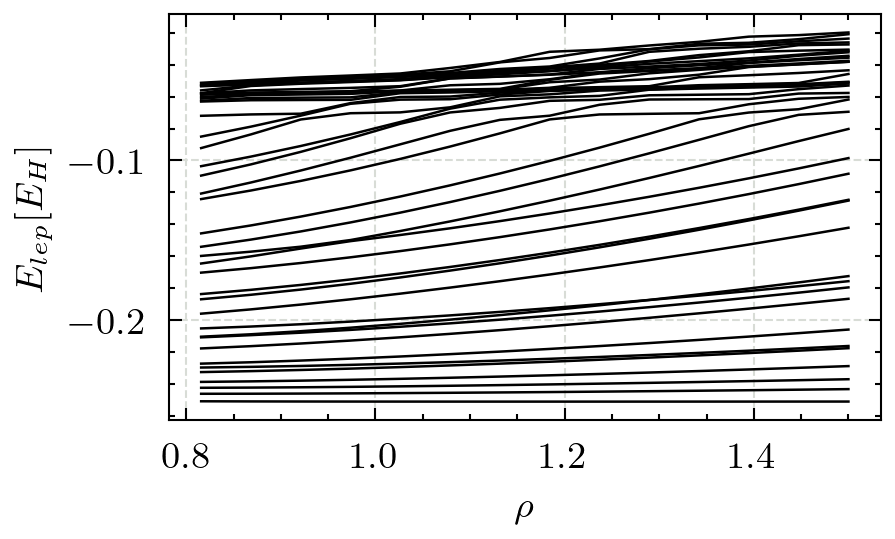}
   \caption{
      Energy of Q-even states at $R=0.8$ as a function of
		a real scaling parameter $\rho$. 
   }
   \label{fig:real_scaling}
\end{figure}

As is well known, continuum states can be discretized by enclosing the 
quantum system into a box with infinite walls. The energies of the 
continuum states increase quadratically with the box size, while 
those of bound states are box-size independent, once the box is 
sufficiently large to fully contain the state. Within the present 
approach, the effect of an enclosing box can be emulated by a 
simultaneous scaling of all exponents of the basis functions, 
i.\,e.\ of the base. More accurately, it is the radial coordinate 
that should be scaled, and thus within the here adopted 
prolate-spheroidal coordinates it is the $\xi$ coordinate.  
Thus the base parameters $u$ and $w$ are multiplied 
with a real scaling parameter $\rho$,
\begin{eqnarray}
   u(\rho) = \rho\,u\text{,}\quad
   w(\rho) = \rho\,w \; .
\end{eqnarray}
Figure \ref{fig:real_scaling} shows the leptonic energies of the  
Q-even $\Sigma$ states at $R=0.8$ as a function 
of $\rho$. The lowest state, the \HHbar\ molecular ground state, 
is basically independent of $\rho$. However, the states with 
slightly higher energy show the quadratic dependence on $\rho$ 
expected for continuum states, in this case states describing a free 
positronium. At higher energies more states with almost constant 
energy appear, corresponding to the excited molecular states 
and showing avoided crossings with the quadratically increasing 
states. Therefore these states indeed describe free positronium 
with additional quanta of kinetic energy.

\begin{figure}
   \includegraphics[width=\columnwidth]{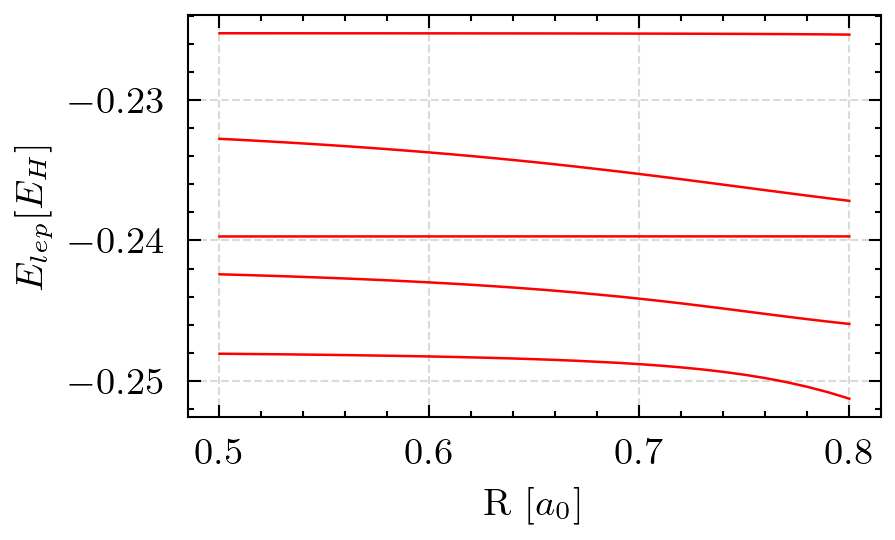}
   \caption{
		Leptonic energies of the ground state and four excited
		states (also Q-even symmetry) for inter-hadronic 
		separations around the critical distance $R_c=0.77\,a_0$.
   }
   \label{fig:qpos_subcritical}
\end{figure}

With the obtained positronium states life-time calculations for the bound
states (more accurately metastable states) should in principle become 
possible, either directly from the real stabilization graph or, even 
better, using the complex-scaling method as was, e.\,g., adopted in 
the four-body calculations in \cite{anti:steg16,anti:yama18b}. 
Of course, this would be especially 
interesting for the leptonic ground state, but no avoided crossing 
of the leptonic ground state with a positronium state can be 
conclusively observed in the region above or below the critical 
distance, see Fig.~\ref{fig:qpos_subcritical}. Again, this would 
be in accordance with the idea that the rearrangement process 
in the ground state is not due to a Feshbach-type resonance, but 
due to a continuous transition from the bound into a continuum 
state. Noteworthy, this seems to be different for 
collisions occurring on excited-state potential curves, since they 
show clear avoided crossings. 

\subsection{H--H vs.\ {$\bar{\mathbf{H}}$}--H collisions}

\begin{figure}
   \includegraphics[width=\columnwidth]{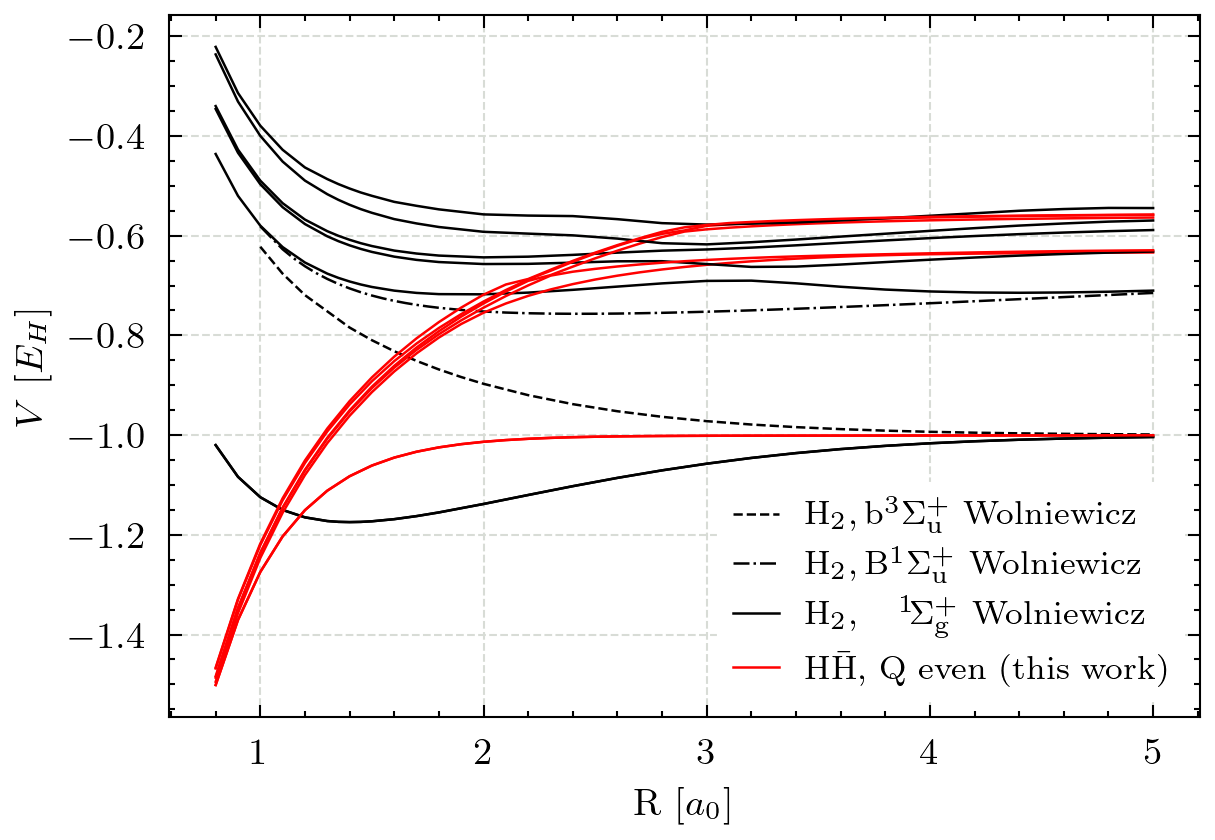}
   \caption{
		Comparison of selected Born-Oppenheimer potential curves ($\Sigma$ symmetry) 
		for $\mathrm{H_2}$ (literature data) and \HHbar\ (red curves, this work). 
		For H$_2$ a selection of lowest lying $X\,^1\Sigma_g^+$ 
		(black solid lines, \cite{dia:woln93}) states as well as the 
		$b\,^3\Sigma_u^+$ (black dashed line, \cite{dia:stas99}) and the 
		$B\,^1\Sigma_u^+$ (black chain line, \cite{dia:stas02}) states 
		are shown.}
   \label{fig:h2_comparison}
\end{figure}
In order to highlight the role of the excited leptonic states of \HHbar\ for 
antihydrogen-hydrogen collisions, it is instructive to compare to the 
collisions between two hydrogen atoms. Figure \ref{fig:h2_comparison} shows 
the lowest lying Q-even potential curves obtained in this work with some of 
the very precisely known potential curves of $\mathrm{H_2}$. Within the here 
adopted non-relativistic approximation, the collision of a 
ground-state H atom with either another H atom or an \Hbar\ atom, both also 
being in their ground states, would start at the same energy: 
$-1.0\,E_H + E_{\rm kin}$ where $E_{\rm kin}$ is the translation energy of 
the two atoms. For small collision energies $E_{\rm kin}$ there is in the 
case of H$_2$ only a single energetically accessible excited state, the 
triplet $b\,^3\Sigma_u^+$, to which the coupling is very small, as it 
involves a spin flip and relativistic effects are very small for light 
atoms. All the other excited states are separated from the ground state by an 
energy gap of some eV and are thus not energetically accessible for small 
collision energies. This is completely different for \HHbar\ where 
in fact for all Born-Oppenheimer excited-state curves one has 
$V\rightarrow -\infty$ for $R\rightarrow 0$. Therefore, in principle, 
\emph{all} excited leptonic states become degenerate with $-1.0\,E_H + E_{\rm kin}$ 
at some inter-hadronic separation $R$, independent of the collision energy 
$E_{\rm kin} \ge 0$. It should be emphasised that here not only the earlier 
considered leptonic states describing the protonium and positronium 
rearrangement channel are meant, but also those "molecular" states that 
correspond asymptotically to one (or both) atoms in an excited state. 
Evidently, a quantitative treatment of possible leptonic excitations needs 
to be performed in accurate scattering calculations of ground-state 
H--\Hbar\ collisions.

%%%\subsection{Rovibronic \HHbar\ States: Resonances}
%

\begin{figure}
   \includegraphics[width=\columnwidth]{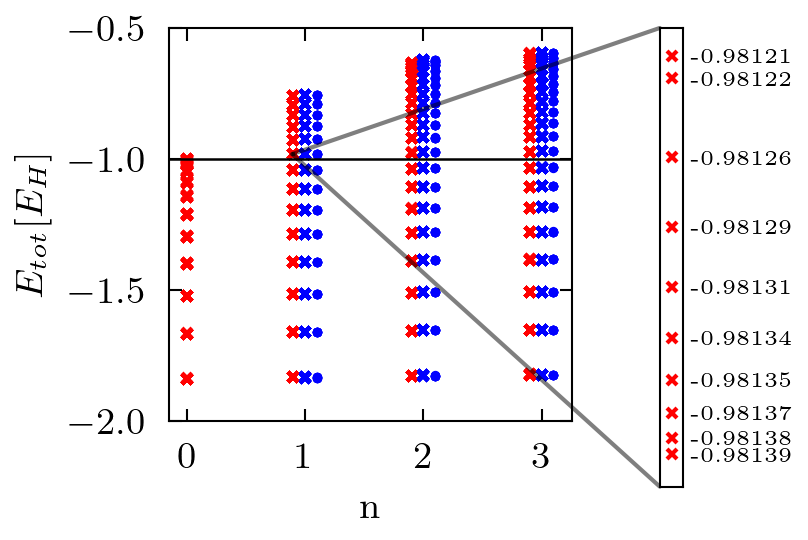}
   \caption{
		Spectrum of Q-even (red) and Q-odd (blue) rovibronic states 
		for rotational quantum numbers $J \in {0,..9}$ and 
		different leptonic states $n$. In the case of Q-odd states 
		the results obtained with two different extrapolations of 
		the potential curves at small interhadronic separations, 
		linear (crosses) or using a 6th-order polynomial (dots),  
		are shown. 
 		The black line marks the dissociation
		energy of \HHbar. Thus the energy range just above this 
		threshold is most relevant for (low-energy) scattering processes 
		of \Hbar\ and H, if both are initially in their ground states. 
		Since the rotational energy spacing is too small to be visible 
		on this scale, the enlarged spectrum in the inset resolves some 
		of the rotational states just above the ground-state 
		dissociation threshold.
	}
   \label{fig:rovib_states_scatter}
\end{figure}

The impact of the excited leptonic states is expected to be especially 
large, if the H and \Hbar\ ground-state atoms in the incoming collision 
channel are energetically degenerate with some rovibrational state of 
a leptonically excited molecular state, i.\,e.\ if a scattering 
resonance occurs. Figure \ref{fig:rovib_states_scatter} 
shows an excerpt of the ro-vibrational spectrum, i.\,e., a selection 
of rovibrational states for the lowest lying leptonic molecular states. 
Evidently, the excited leptonic states support ro-vibrational
states in the region just above the dissociation threshold, that should 
allow for the occurrence of scattering resonances in H--\Hbar-collisions, 
even for very low collision energies. Considering the plethora of 
excited leptonic states, a small fraction been shown in 
Fig.~\ref{fig:bo_overview}, it is evident that a large number of 
resonance occurs, even if only the Q-even states are considered, 
since the Q-odd states should for symmetry reasons not couple to 
the Q-even ground state, if CP symmetry is conserved. Note, that 
for sufficiently large collision energies to overcome the centrifugal 
barrier, also Q-even states of symmetries different from $\Sigma$, 
$\Pi,\Delta,\dots$ are expected to lead to resonances in 
ground-state H--\Hbar\ collisions, as also these potential curves 
tend to $-\infty$ for $R\rightarrow 0$. The large 
density of resonances is, of course, counteracted by the fact  
that the leptonic coupling matrix elements between the ground state 
and an excited leptonic state $n$ decreases with increasing $n$ due 
to the increasing number of nodes of the final-state wave functions. 
On the other hand, since all molecular \HHbar\ states are metastable 
due to the open rearrangement channel into protonium and positronium 
as well as the possibility of leptonic and hadronic annihilation or 
dipole-allowed radiative relaxation, the rovibronic states  
possess a finite lifetime and width. Thus it is expected that 
the resonance condition is fulfilled at practically any collision 
energy.

While the main finding of the present work is the occurrence of 
a high density of resonances in ground-state H--\Hbar\ collisions, 
but not their exact positions, the uncertainties due to the 
treatment (extrapolation) below the critical distance may shed 
some doubt on the findings. The rovibronic ground state of \HHbar\ 
obtained with the present approach yields an energy of \au{-460.347},
which is very close to the ground state energy of positronium plus
protonium of \au{-459.288}. This is expected, since the Born-Oppenheimer 
potential curves at small inter-hadronic separations are dominated by 
the $1/R$ term and thus the details of the leptonic energies are of 
minor importance. Therefore, different extrapolations resulted in very 
similar energies. Nevertheless, in order to obtain an independent 
cross-check of the present Born-Oppenheimer treatment (including the 
extrapolations described above), a full non-relativistic  
four-body calculation of the ground-state energy of the 
\HHbar\ molecule was performed using the code 
ATOM-MOL-nonBO \cite{gen:bubi20}. The energy obtained
from this ansatz is \au{-459.219} and the relative deviation 
of the present Born-Oppenheimer calculation is $2.4\times 10^{-3}$. 
Thus both treatments are in quite good quantitative agreement, despite 
the uncertainties from adopting the Born-Oppenheimer approximation 
and extrapolating the Born-Oppenheimer potential curves. Noteworthy, 
however, the four-body code yields a ground-state energy above the 
positronium-protonium ground state which is in accordance with the 
existence of a critical distance, while the present Born-Oppenheimer 
result would falsely suggest the molecule to be stable against 
rearrangement into positronium and protonium.

\section{Conclusion}
The $\Sigma$ spectrum of the \HHbar\ molecule was calculated
using \KW\ basis functions and the Born-Oppenheimer approximation. 
Adopting a dual-base basis set, i.\,e.\ \KW\ basis sets with two 
different sets of non-integer exponents, it was possible to 
reveal free positronium states which repeatedly cross the leptonic bound
states, except for the ground and first excited states, leading to a 
large number of avoided crossings. The ground state (Q even) and the 
lowest lying excited states (Q even and Q odd) seem to change their characters 
around $R=0.8a_0$ and $R=2.0a_0$, respectively. However, no conclusive 
avoided crossing with positronium states was found. Whether the change of 
character is in these cases a continous change of state character or if 
the avoided crossing is not visible due to a deficiency of the adopted 
basis sets needs further investigation. Extrapolating the leptonic potential 
curves to inter-hadronic distances below the critical distance, the 
ro-vibronic spectrum was calculated for a number of leptonic states. 
This calculation is not meant to be of high accuracy, since part of the 
Born-Oppenheimer potentials were extrapolated and the effects of the 
strong-force interaction completely ignored. A full four-body calculation 
of the rovibronic ground state confirmed that the here obtained 
Born-Oppenheimer solutions appear to be reasonably good, despite the 
earlier reported break-down of the Born-Oppenheimer approximation at the 
critical distance at which the energy of a positronium-protonium 
system starts to lie below the one of the \HHbar\ molecule. Thus it 
can be concluded that besides the already earlier known need for 
including explicitly the positronium-protonium rearrangement channel 
into scattering calculations describing H--\Hbar\ collisions, also the 
excited leptonic states and corresponding resonances need to be considered. 
The present work is expected to provide a good starting point for such 
scattering calculations.

\section{Acknowledgements}
J.S. acknowledges financial support by the German Federal Ministry of 
Research, Technology and Space (BMFTR) within ErUM-Pro 05A23PMA.

\appendix
\FloatBarrier
\section{Symmetry-Adapted \HHbar\ Basis Functions}\label{app:filter_rules}
Besides the linear dependencies due to the finite numerical precision 
(which is prevented using variable-precision arithmetics in the in this 
work adopted modified H2SOLV code \cite{dia:pach16}), two symmetry-adapted 
\KW-type basis functions can become identical (and thus linear dependent) 
or zero valued (and thus inappropriate) for specific choices of the 
basis-set parameters. While this problem does not occur for the final 
bases used in this work for producing the shown results, it can 
occur during the basis-set optimization in which two of the base 
parameters may become equal. This problem has been circumvented by 
removing those basis functions from the basis set that are zero valued 
(or would occur twice). In this section the construction of 
symmetry-adapted basis functions avoiding/removing the redundant ones 
is described. The \KW\ functions are versatile and comprise four well 
known types of basis sets depending on the symmetry of the base parameters, 
see Table \ref{tab:base_symmetries} \cite{dia:pach16}. In the following, 
the names of these basis sets will be used for better readability.

\begin{table}
   \caption{Base Symmetries.}
   \begin{tabular}{c|c}
      Basis Set & Base\\\hline
      James-Coolidge Symmetric   & $y=x=0 \land u=w$\\\hline
      James-Coolidge Asymmetric  & $y=x=0 \land u\neq w$\\\hline
      $\mathrm{H}_2$             & $y=-x=u=w$\\\hline
      $\mathrm{H}^-\mathrm{H}^+$ & $y=x=u=w$\\
   \end{tabular}
   \label{tab:base_symmetries}
\end{table}

The symmetry adapted linear combination in equation \eqref{eq:salc}
\begin{align}
   \phi_\mathbf{n}^{\pm}& \propto \exp \left (-y\eta_1 - x\eta_2 - u\xi_1 - w\xi_2\right)\,\eta_1^{n_1}\,\eta_2^{n_2}\,\xi_1^{n_3}\,\xi_2^{n_4}\nonumber\\
                      &\pm (-1)^{n_1 + n_2} \exp \left (+y\eta_2 + x\eta_1 - u\xi_2 - w\xi_1\right)\,\eta_2^{n_1}\,\eta_1^{n_2}\,\xi_2^{n_3}\,\xi_1^{n_4}
   \label{eq:salc_full}
\end{align}
is here rewritten, but omitting the factors of $r_{e\bar{e}}$ and $R$ as they
are invariant under the Q-symmetry operation. Whether a basis function is 
zero-valued can be seen directly from equation \eqref{eq:salc_full}, once 
a symmetry-adpated base is assumed.
To determine whether a basis function might lead to linear dependencies
in the overlap matrix, i.e.\ the brakets of
$\phi_\mathbf{n}$ with $\phi_\mathbf{m}^\pm$ and
$\phi^\pm_{\sigma(\mathbf{m})}$
with $\sigma \in S_4$ have to be inspected.
If for a fixed $\phi_\mathbf{m}$ two or more permutations $\sigma_1, \sigma_2,\dots$
exist such that ${\braket{\phi|\phi_{\sigma_1(\mathbf{m})}^\pm} = \braket{\phi|\phi_{\sigma_2(\mathbf{m})}^\pm}}$
and $ \phi_{\sigma_1(\mathbf{m})}^\pm, \phi_{\sigma_2(\mathbf{m})}^\pm$ are both contained in the
basis, the overlap matrix will contain identical rows.

\subsection{James-Coolidge Symmetric}
With $y=x=0 \land u=w$ equation \eqref{eq:salc_full} simplifies to
\begin{align*}
   \phi_\mathbf{m}^{\pm} \propto &\exp \left (-u(\xi_1 - \xi_2)\right)\,[ \eta_1^{m_1}\,\eta_2^{m_2}\,\xi_1^{m_3}\,\xi_2^{m_4}\\
                      &\pm (-1)^{m_1 + m_2}\,\eta_2^{m_1}\,\eta_1^{m_2}\,\xi_2^{m_3}\,\xi_1^{m_4} ]
   \label{eq:james_coolidge}
\end{align*}
and evaluates to zero for $m_1=m_2 \land m_3=m_4$ and Q-odd symmetry.
Linear dependencies, as being caused by  
\begin{align*}
   \braket{\phi_\mathbf{n}|\phi_\mathbf{m}^\pm} = &\exp(-2u(\xi_1 + \xi_2))\\
       &[\eta_1^{n_1 + m_1}\,\eta_2^{n_2 + m_2}\,\xi_1^{n_3 + m_3}\,\xi_2^{n_4 + m_4}\\
       &\pm(-1)^{m_1 + m_2} \eta_1^{n_1 + m_2}\,\eta_2^{n_2 + m_1}\,\xi_1^{n_3 + m_4}\,\xi_2^{n_4 + m_3}]\quad\text{,}
\end{align*}
occur, if $m_1, m_2$ and $m_3, m_4$ are permuted respectively.
To obtain a valid basis it has to be restricted according to 
\[m_1 > m_3 \lor (m_1=m_3 \land m_3 > m_4)\]

\subsection{James-Coolidge Asymmetric}
With $y=x=0 \land u=w$ equation \eqref{eq:salc_full} simplifies to
\begin{align*}
   \phi_{\mathbf{m}}^{\pm} \propto &\exp \left (-u\xi_1 -w\xi_2\right)\, \eta_1^{m_1}\,\eta_2^{m_2}\,\xi_1^{m_3}\,\xi_2^{m_4}\\
                      &\pm (-1)^{m_1 + m_2}\exp \left (-w\xi_1 - u\xi_2)\right)\,\eta_2^{m_1}\,\eta_1^{m_2}\,\xi_2^{m_3}\,\xi_1^{m_4} 
\end{align*}
which will not evaluate to zero.
Linear dependencies will not occur as well, as the exponents in
\begin{align*}
   &\braket{\phi_\mathbf{n}|\phi_\mathbf{m}^\pm} = \exp(-2u\xi_1 - 2w\xi_2)\eta_1^{n_1 + m_1}\,\eta_2^{n_2 + m_2}\,\xi_1^{n_3 + m_3}\,\xi_2^{n_4 + m_4}\\
   &\pm s\exp(-(u+w)(\xi_1 + \xi_2))\eta_1^{n_1 + m_2}\,\eta_2^{n_2 + m_1}\,\xi_1^{n_3 + m_4}\,\xi_2^{n_4 + m_3}\quad\text{,}
\end{align*}
where $s = (-1)^{m_1 + m_2}$, differ.

\subsection{$\mathrm{H_2}$}
With $y=-x=u=w$ equation \eqref{eq:salc_full} simplifies to
\begin{align*}
   \phi_\mathbf{m}^\pm \propto &\exp(y\eta_1 - y\eta_2 - y\xi_1 - y\xi_2) [ \eta_1^{m_1}\,\eta_2^{m_2}\,\xi_1^{m_3}\,\xi_2^{m_4}\\
   &\pm (-1)^{m_1 + m_2}\,\eta_2^{m_1}\,\eta_1^{m_2}\,\xi_2^{m_3}\,\xi_1^{m_4} ]
\end{align*}
which again will simplify to zero for $m_1 = m_2 \land m_3 = m_4$.
Linear dependencies arise due to
\begin{align*}
   \braket{\phi_\mathbf{n}|\phi_\mathbf{m}^\pm} = &\exp(2y(\eta_1 - \eta_2 - \xi_1 - \xi_2)) \\
       &[ \eta_1^{n_1 + m_1}\,\eta_2^{n_2 + m_2}\,\xi_1^{n_3 + m_3}\,\xi_2^{n_4 + m_4}\\
       &\pm(-1)^{m_1 + m_2} \eta_1^{n_1 + m_2}\,\eta_2^{n_2 + m_1}\,\xi_1^{n_3 + m_4}\,\xi_2^{n_4 + m_3} ]\quad\text{,}
\end{align*}
for permutations of $m_1, m_2$ and $m_3, m_4$. The same
restrictions as for the symmetric James-Coolidge basis have to be applied.

\subsection{$\mathrm{H}^-\mathrm{H}^+$}
With $y=x=u=w$ equation \eqref{eq:salc_full} simplifies to
\begin{align*}
   \phi_\mathbf{m}^\pm \propto &\exp(-y(\eta_1 + \eta_2 + \xi_1 + \xi_2))\,\eta_1^{m_1}\,\eta_2^{m_2}\,\xi_1^{m_3}\,\xi_2^{m_4}\\
   &\pm(-1)^{m_1 + m_2}\exp(-y(-\eta_1 - \eta_2 + \xi_1 + \xi_2)) \eta_2^{m_1}\,\eta_1^{m_2}\,\xi_2^{m_3}\,\xi_1^{m_4}
\end{align*}
which again, due to the different exponential dampening factors, will
neither evaluate to zero, nor yield linear dependencies.

\section{Born-Oppenheimer potential curves}\label{sec:BO_curves}
\begin{figure*}
\includegraphics[width=\textwidth]{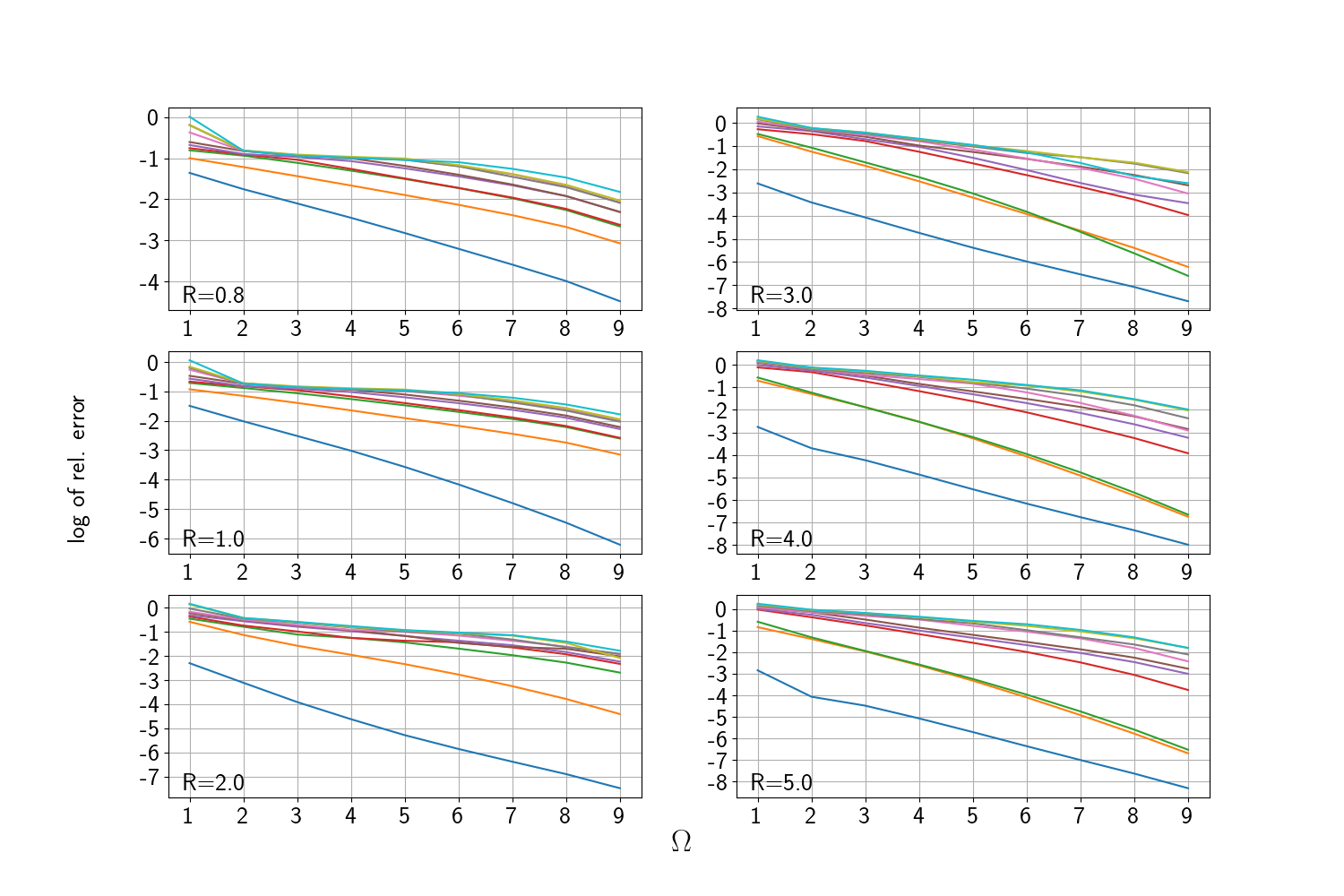}
\caption{
	Relative differences of the Born-Oppenheimer potential curves
	for Q-positive states at different inter-hadronic separations
	to a reference calculation ($\Omega=10$).
	The relative difference decreases exponentially as a function of $\Omega$.
	The different colors amount to different excited states. In increasing order:
	blue, orange, green, red, purple, brown, pink, gray, olive, and cyan.
}
\label{app:convergence1}
\end{figure*}

\begin{figure*}
\includegraphics[width=\textwidth]{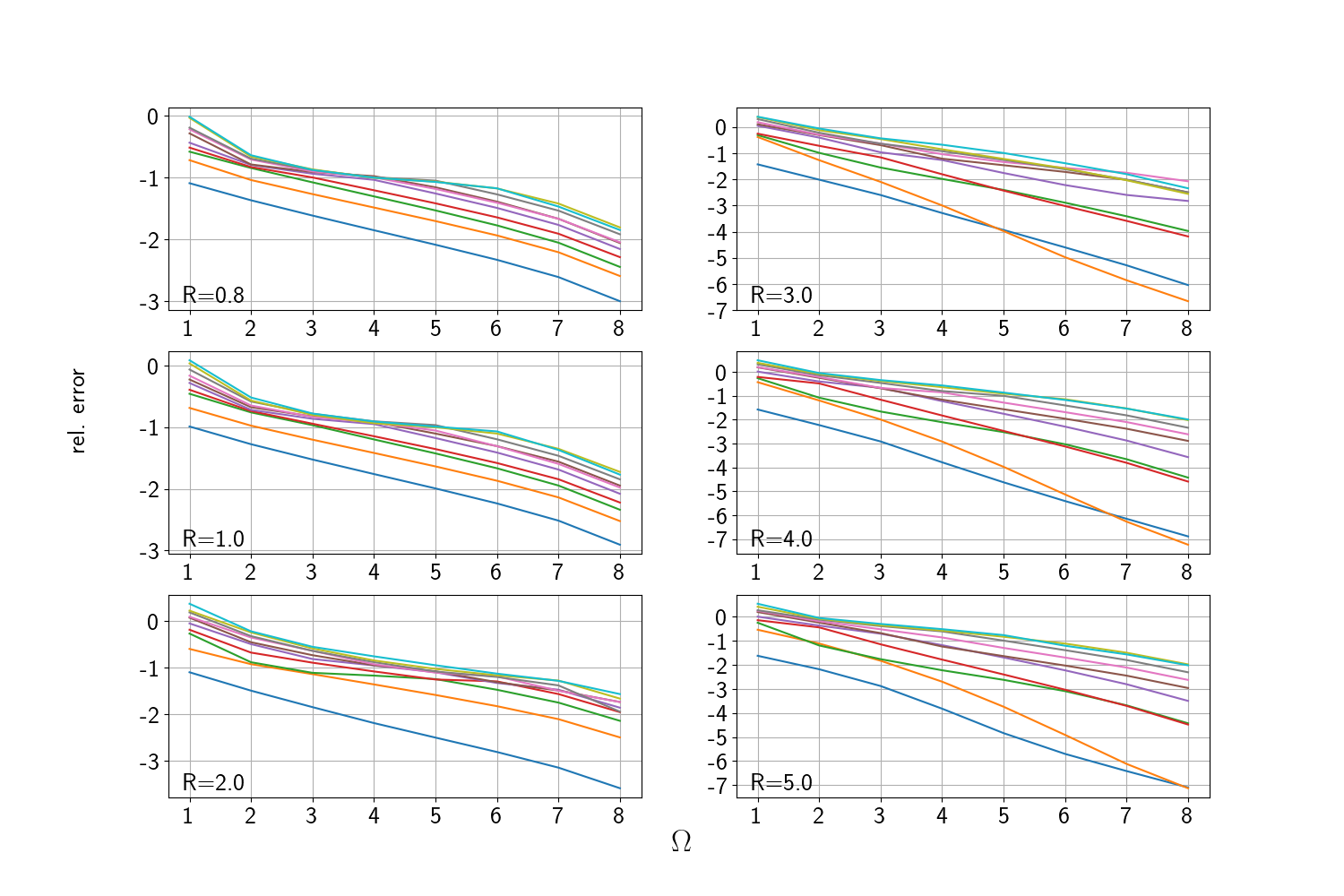}
\caption{
    as Fig.~\ref{app:convergence1}, but for the ten lowest lying 
    Q-odd instead of Q-even states. 
}
\label{app:convergence2}
\end{figure*}

The numerical values of the 10 energetically lowest-lying $\Sigma$ states 
of \HHbar\ are given in numerical form, both for Q-even and Q-odd symmetries, 
in Tables \ref{tab:bo_qpos} and \ref{tab:bo_qneg}, respectively. The 
corresponding convergences studies as a function of the basis-set 
parameter $\Omega$ are shown in Figs.~\ref{app:convergence1} and 
\ref{app:convergence2}, respectively.

\begin{table*}
\caption{
   Born-Oppenheimer potential curves (in Hartree) for the 10 lowest lying $\Sigma$ states 
   ($n=0$ to $n=9$) with Q-even symmetry. (The inter-hadronic separations $R$ are given 
   in Bohr units $a_0$.)
}
\begin{tabular}{c|c|c|c|c|c|c|c|c|c|c}
R & $n = 0$ & $n = 1$ & $n = 2$ & $n = 3$ & $n = 4$ & $n = 5$ & $n = 6$ & $n = 7$ & $n = 8$ & $n = 9$\\
\hline
0.8 & -1.501279 & -1.495951 & -1.489710 & -1.487174 & -1.475321 & -1.473540 & -1.467338 & -1.454995 & -1.453036 & -1.436223 \\
1.0 & -1.274441 & -1.247239 & -1.239823 & -1.239344 & -1.226919 & -1.225048 & -1.217280 & -1.208463 & -1.203493 & -1.186224 \\
1.1 & -1.203236 & -1.156483 & -1.149193 & -1.148631 & -1.136884 & -1.134174 & -1.126333 & -1.118905 & -1.112659 & -1.095441 \\
1.2 & -1.150521 & -1.080825 & -1.073741 & -1.072903 & -1.061828 & -1.058426 & -1.050535 & -1.044307 & -1.036973 & -1.020691 \\
1.3 & -1.111461 & -1.016808 & -1.009928 & -1.008800 & -0.998371 & -0.994327 & -0.986390 & -0.981301 & -0.972951 & -0.958115 \\
1.4 & -1.082505 & -0.961953 & -0.955299 & -0.953845 & -0.944105 & -0.939384 & -0.931403 & -0.927491 & -0.918106 & -0.904780 \\
1.5 & -1.061035 & -0.914453 & -0.908077 & -0.906212 & -0.897283 & -0.891769 & -0.883755 & -0.881111 & -0.870617 & -0.858846 \\
1.6 & -1.045120 & -0.872970 & -0.866967 & -0.864529 & -0.856606 & -0.850112 & -0.842126 & -0.840765 & -0.829125 & -0.818906 \\
1.7 & -1.033332 & -0.836547 & -0.831031 & -0.827746 & -0.821060 & -0.813364 & -0.805966 & -0.804856 & -0.792604 & -0.783823 \\
1.8 & -1.024610 & -0.804630 & -0.799524 & -0.795046 & -0.789741 & -0.780712 & -0.774625 & -0.772268 & -0.760300 & -0.752687 \\
1.9 & -1.018170 & -0.777261 & -0.771516 & -0.765784 & -0.761764 & -0.751521 & -0.746661 & -0.742999 & -0.731869 & -0.724895 \\
2.0 & -1.013428 & -0.754619 & -0.745980 & -0.739446 & -0.736406 & -0.725302 & -0.721376 & -0.716654 & -0.709675 & -0.702571 \\
2.1 & -1.009947 & -0.736071 & -0.722550 & -0.715615 & -0.713240 & -0.701892 & -0.698427 & -0.696293 & -0.692602 & -0.680558 \\
2.4 & -1.004216 & -0.697160 & -0.672293 & -0.663384 & -0.655974 & -0.654580 & -0.642020 & -0.640141 & -0.633018 & -0.624937 \\
2.5 & -1.003256 & -0.688019 & -0.666591 & -0.646754 & -0.639284 & -0.638048 & -0.625665 & -0.623908 & -0.616519 & -0.612151 \\
2.6 & -1.002572 & -0.680219 & -0.661780 & -0.631393 & -0.623879 & -0.622788 & -0.611472 & -0.609056 & -0.603059 & -0.599830 \\
2.7 & -1.002088 & -0.673531 & -0.657699 & -0.617167 & -0.609655 & -0.608744 & -0.601313 & -0.595188 & -0.591683 & -0.590968 \\
2.8 & -1.001749 & -0.667769 & -0.654224 & -0.603982 & -0.597385 & -0.596039 & -0.593433 & -0.586953 & -0.583880 & -0.581767 \\
2.9 & -1.001512 & -0.662789 & -0.651255 & -0.592280 & -0.590564 & -0.583958 & -0.583413 & -0.582345 & -0.580259 & -0.569698 \\
3.0 & -1.001348 & -0.658470 & -0.648709 & -0.587360 & -0.580452 & -0.579750 & -0.577602 & -0.572287 & -0.571010 & -0.558339 \\
3.2 & -1.001153 & -0.651444 & -0.644633 & -0.581347 & -0.575057 & -0.573029 & -0.559117 & -0.552510 & -0.551414 & -0.550236 \\
3.3 & -1.001093 & -0.648592 & -0.642999 & -0.579023 & -0.572960 & -0.571152 & -0.550433 & -0.549389 & -0.547967 & -0.546905 \\
3.4 & -1.001048 & -0.646106 & -0.641576 & -0.577021 & -0.571124 & -0.569511 & -0.548360 & -0.546172 & -0.545023 & -0.541351 \\
3.5 & -1.001010 & -0.643943 & -0.640328 & -0.575283 & -0.569510 & -0.568072 & -0.546731 & -0.544562 & -0.543330 & -0.539134 \\
3.6 & -1.000976 & -0.642071 & -0.639218 & -0.573765 & -0.568089 & -0.566806 & -0.545311 & -0.543126 & -0.541794 & -0.537082 \\
3.7 & -1.000943 & -0.640474 & -0.638206 & -0.572431 & -0.566832 & -0.565690 & -0.544060 & -0.541840 & -0.540387 & -0.535171 \\
3.8 & -1.000911 & -0.639144 & -0.637248 & -0.571254 & -0.565719 & -0.564705 & -0.542952 & -0.540681 & -0.539084 & -0.533386 \\
3.9 & -1.000878 & -0.638077 & -0.636307 & -0.570211 & -0.564728 & -0.563835 & -0.541966 & -0.539631 & -0.537867 & -0.531720 \\
4.0 & -1.000843 & -0.637241 & -0.635378 & -0.569283 & -0.563845 & -0.563065 & -0.541085 & -0.538675 & -0.536721 & -0.530169 \\
4.1 & -1.000807 & -0.636579 & -0.634486 & -0.568453 & -0.563055 & -0.562383 & -0.540294 & -0.537800 & -0.535632 & -0.528730 \\
4.2 & -1.000770 & -0.636040 & -0.633653 & -0.567709 & -0.562346 & -0.561778 & -0.539580 & -0.536996 & -0.534591 & -0.527403 \\
4.3 & -1.000733 & -0.635589 & -0.632892 & -0.567037 & -0.561709 & -0.561241 & -0.538933 & -0.536254 & -0.533590 & -0.526192 \\
4.4 & -1.000694 & -0.635205 & -0.632203 & -0.566430 & -0.561134 & -0.560764 & -0.538343 & -0.535566 & -0.532623 & -0.525101 \\
4.5 & -1.000656 & -0.634871 & -0.631581 & -0.565878 & -0.560615 & -0.560339 & -0.537801 & -0.534927 & -0.531686 & -0.524127 \\
4.6 & -1.000618 & -0.634579 & -0.631023 & -0.565374 & -0.560144 & -0.559960 & -0.537300 & -0.534331 & -0.530776 & -0.523251 \\
4.7 & -1.000580 & -0.634319 & -0.630521 & -0.564912 & -0.559717 & -0.559622 & -0.536831 & -0.533775 & -0.529892 & -0.522443 \\
4.8 & -1.000543 & -0.634086 & -0.630070 & -0.564487 & -0.559333 & -0.559316 & -0.536390 & -0.533253 & -0.529034 & -0.521675 \\
4.9 & -1.000507 & -0.633874 & -0.629664 & -0.564094 & -0.559050 & -0.558974 & -0.535970 & -0.532765 & -0.528203 & -0.520928 \\
5.0 & -1.000473 & -0.633679 & -0.629299 & -0.563730 & -0.558807 & -0.558651 & -0.535566 & -0.532306 & -0.527399 & -0.520192 \\
\end{tabular}
\label{tab:bo_qpos}
\end{table*}

\begin{table*}
\caption{
   As Table \ref{tab:bo_qpos}, but for states with Q-odd symmetry.
}
\begin{tabular}{c|c|c|c|c|c|c|c|c|c|c}
R & $n = 0$ & $n = 1$ & $n = 2$ & $n = 3$ & $n = 4$ & $n = 5$ & $n = 6$ & $n = 7$ & $n = 8$ & $n = 9$\\
\hline
0.8 & -1.495716 & -1.488765 & -1.486060 & -1.476993 & -1.473624 & -1.468267 & -1.460019 & -1.453327 & -1.446747 & -1.439091 \\
1.0 & -1.245597 & -1.238857 & -1.235332 & -1.227564 & -1.223610 & -1.216239 & -1.210941 & -1.204735 & -1.196119 & -1.188518 \\
1.1 & -1.154622 & -1.147960 & -1.144035 & -1.136869 & -1.132599 & -1.124268 & -1.120474 & -1.114345 & -1.104746 & -1.098180 \\
1.2 & -1.078798 & -1.072206 & -1.067882 & -1.061296 & -1.056718 & -1.047468 & -1.045117 & -1.039000 & -1.028487 & -1.022983 \\
1.3 & -1.014628 & -1.008112 & -1.003387 & -0.997374 & -0.992504 & -0.982581 & -0.981185 & -0.975235 & -0.963994 & -0.959339 \\
1.4 & -0.959619 & -0.953195 & -0.948072 & -0.942637 & -0.937502 & -0.927719 & -0.925656 & -0.920574 & -0.908979 & -0.904580 \\
1.5 & -0.911943 & -0.905648 & -0.900147 & -0.895304 & -0.889940 & -0.880732 & -0.877144 & -0.873155 & -0.861721 & -0.856751 \\
1.6 & -0.870235 & -0.864152 & -0.858334 & -0.854058 & -0.848524 & -0.839905 & -0.834757 & -0.831449 & -0.820693 & -0.814573 \\
1.7 & -0.833468 & -0.827792 & -0.821830 & -0.817823 & -0.812239 & -0.803979 & -0.797643 & -0.794206 & -0.784595 & -0.777203 \\
1.8 & -0.800908 & -0.796116 & -0.790128 & -0.785501 & -0.780128 & -0.771894 & -0.764891 & -0.760664 & -0.752445 & -0.744008 \\
1.9 & -0.772410 & -0.768889 & -0.762087 & -0.756250 & -0.751333 & -0.742911 & -0.735656 & -0.730378 & -0.723539 & -0.714451 \\
2.0 & -0.749520 & -0.743916 & -0.736460 & -0.729691 & -0.725280 & -0.716583 & -0.709346 & -0.702963 & -0.697396 & -0.688072 \\
2.1 & -0.731616 & -0.720372 & -0.712948 & -0.705540 & -0.701599 & -0.692601 & -0.685561 & -0.678156 & -0.675376 & -0.672553 \\
2.2 & -0.716923 & -0.698758 & -0.691427 & -0.683520 & -0.680017 & -0.670816 & -0.666444 & -0.663609 & -0.655333 & -0.651448 \\
2.3 & -0.704631 & -0.678966 & -0.671724 & -0.663434 & -0.660681 & -0.658906 & -0.650450 & -0.644104 & -0.634835 & -0.631610 \\
2.4 & -0.694241 & -0.660805 & -0.654263 & -0.652928 & -0.644843 & -0.642029 & -0.632155 & -0.626150 & -0.617227 & -0.613529 \\
2.5 & -0.685396 & -0.648701 & -0.644034 & -0.636891 & -0.627915 & -0.625399 & -0.615640 & -0.609892 & -0.604815 & -0.600794 \\
2.6 & -0.677824 & -0.644516 & -0.628606 & -0.621582 & -0.612506 & -0.610090 & -0.602250 & -0.596269 & -0.595551 & -0.592407 \\
2.7 & -0.671311 & -0.641024 & -0.614320 & -0.607564 & -0.599509 & -0.596059 & -0.592868 & -0.590285 & -0.583718 & -0.582637 \\
2.8 & -0.665688 & -0.638082 & -0.601121 & -0.595540 & -0.590677 & -0.586563 & -0.582986 & -0.581067 & -0.579965 & -0.569714 \\
2.9 & -0.660816 & -0.635595 & -0.589980 & -0.587512 & -0.583026 & -0.580275 & -0.576920 & -0.570715 & -0.569063 & -0.560288 \\
3.0 & -0.656583 & -0.633486 & -0.585221 & -0.579760 & -0.576711 & -0.574246 & -0.569273 & -0.559738 & -0.557746 & -0.555656 \\
3.1 & -0.652896 & -0.631693 & -0.582075 & -0.576958 & -0.571918 & -0.565992 & -0.558810 & -0.554147 & -0.552363 & -0.551810 \\
3.2 & -0.649679 & -0.630166 & -0.579495 & -0.574555 & -0.569892 & -0.555915 & -0.552071 & -0.550142 & -0.549524 & -0.548094 \\
3.3 & -0.646867 & -0.628864 & -0.577303 & -0.572467 & -0.568134 & -0.549862 & -0.548145 & -0.547521 & -0.546223 & -0.544523 \\
3.4 & -0.644405 & -0.627751 & -0.575411 & -0.570644 & -0.566603 & -0.547957 & -0.546383 & -0.545771 & -0.542881 & -0.537319 \\
3.5 & -0.642248 & -0.626802 & -0.573765 & -0.569048 & -0.565267 & -0.546352 & -0.544855 & -0.544235 & -0.541434 & -0.531106 \\
3.6 & -0.640356 & -0.625990 & -0.572324 & -0.567648 & -0.564099 & -0.544956 & -0.543512 & -0.542881 & -0.540159 & -0.529711 \\
3.7 & -0.638695 & -0.625298 & -0.571055 & -0.566416 & -0.563077 & -0.543730 & -0.542326 & -0.541684 & -0.539031 & -0.528507 \\
3.8 & -0.637236 & -0.624709 & -0.569934 & -0.565330 & -0.562180 & -0.542648 & -0.541276 & -0.540624 & -0.538031 & -0.527447 \\
3.9 & -0.635954 & -0.624209 & -0.568939 & -0.564371 & -0.561394 & -0.541688 & -0.540345 & -0.539683 & -0.537143 & -0.526510 \\
4.0 & -0.634828 & -0.623786 & -0.568053 & -0.563522 & -0.560703 & -0.540835 & -0.539517 & -0.538845 & -0.536353 & -0.525677 \\
4.1 & -0.633837 & -0.623432 & -0.567261 & -0.562769 & -0.560097 & -0.540073 & -0.538778 & -0.538097 & -0.535648 & -0.524935 \\
4.2 & -0.632964 & -0.623136 & -0.566552 & -0.562101 & -0.559564 & -0.539390 & -0.538118 & -0.537428 & -0.535017 & -0.524272 \\
4.3 & -0.632197 & -0.622893 & -0.565914 & -0.561506 & -0.559095 & -0.538778 & -0.537528 & -0.536828 & -0.534452 & -0.523678 \\
4.4 & -0.631520 & -0.622696 & -0.565339 & -0.560975 & -0.558683 & -0.538226 & -0.536998 & -0.536289 & -0.533945 & -0.523144 \\
4.5 & -0.630924 & -0.622540 & -0.564818 & -0.560502 & -0.558321 & -0.537728 & -0.536522 & -0.535804 & -0.533489 & -0.522663 \\
4.6 & -0.630398 & -0.622418 & -0.564345 & -0.560078 & -0.558002 & -0.537278 & -0.536093 & -0.535366 & -0.533079 & -0.522229 \\
4.7 & -0.629933 & -0.622328 & -0.563915 & -0.559698 & -0.557722 & -0.536869 & -0.535707 & -0.534970 & -0.532708 & -0.521836 \\
4.8 & -0.629523 & -0.622266 & -0.563522 & -0.559357 & -0.557476 & -0.536496 & -0.535358 & -0.534611 & -0.532374 & -0.521480 \\
4.9 & -0.629160 & -0.622227 & -0.563162 & -0.559051 & -0.557260 & -0.536157 & -0.535042 & -0.534285 & -0.532071 & -0.521157 \\
5.0 & -0.628839 & -0.622209 & -0.562832 & -0.558775 & -0.557070 & -0.535846 & -0.534756 & -0.533989 & -0.531796 & -0.520862 \\
\end{tabular}
\label{tab:bo_qneg}
\end{table*}

\FloatBarrier
%%%
%%%\bibliography{journals,anti,dia,cold,gen}

\begin{thebibliography}{54}%
\makeatletter
\providecommand \@ifxundefined [1]{%
 \@ifx{#1\undefined}
}%
\providecommand \@ifnum [1]{%
 \ifnum #1\expandafter \@firstoftwo
 \else \expandafter \@secondoftwo
 \fi
}%
\providecommand \@ifx [1]{%
 \ifx #1\expandafter \@firstoftwo
 \else \expandafter \@secondoftwo
 \fi
}%
\providecommand \natexlab [1]{#1}%
\providecommand \enquote  [1]{``#1''}%
\providecommand \bibnamefont  [1]{#1}%
\providecommand \bibfnamefont [1]{#1}%
\providecommand \citenamefont [1]{#1}%
\providecommand \href@noop [0]{\@secondoftwo}%
\providecommand \href [0]{\begingroup \@sanitize@url \@href}%
\providecommand \@href[1]{\@@startlink{#1}\@@href}%
\providecommand \@@href[1]{\endgroup#1\@@endlink}%
\providecommand \@sanitize@url [0]{\catcode `\\12\catcode `\$12\catcode
  `\&12\catcode `\#12\catcode `\^12\catcode `\_12\catcode `\%12\relax}%
\providecommand \@@startlink[1]{}%
\providecommand \@@endlink[0]{}%
\providecommand \url  [0]{\begingroup\@sanitize@url \@url }%
\providecommand \@url [1]{\endgroup\@href {#1}{\urlprefix }}%
\providecommand \urlprefix  [0]{URL }%
\providecommand \Eprint [0]{\href }%
\providecommand \doibase [0]{https://doi.org/}%
\providecommand \selectlanguage [0]{\@gobble}%
\providecommand \bibinfo  [0]{\@secondoftwo}%
\providecommand \bibfield  [0]{\@secondoftwo}%
\providecommand \translation [1]{[#1]}%
\providecommand \BibitemOpen [0]{}%
\providecommand \bibitemStop [0]{}%
\providecommand \bibitemNoStop [0]{.\EOS\space}%
\providecommand \EOS [0]{\spacefactor3000\relax}%
\providecommand \BibitemShut  [1]{\csname bibitem#1\endcsname}%
\let\auto@bib@innerbib\@empty
%</preamble>
\bibitem [{\citenamefont {Baur}\ \emph {et~al.}(1996)\citenamefont {Baur},
  \citenamefont {Boero}, \citenamefont {Brauksiepe}, \citenamefont {Buzzo},
  \citenamefont {Eyrich}, \citenamefont {Geyer}, \citenamefont {Grzonka},
  \citenamefont {Hauffe}, \citenamefont {Kilian}, \citenamefont {LoVetere},
  \citenamefont {Macri}, \citenamefont {Moosburger}, \citenamefont {Nellen},
  \citenamefont {Oelert}, \citenamefont {Passaggio}, \citenamefont {Pozzo},
  \citenamefont {R{\"o}hrich}, \citenamefont {Sachs}, \citenamefont {Schepers},
  \citenamefont {Sefzick}, \citenamefont {Simon}, \citenamefont {Stratmann},
  \citenamefont {Stinzing},\ and\ \citenamefont {Wolke}}]{anti:baur96}%
  \BibitemOpen
  \bibfield  {author} {\bibinfo {author} {\bibfnamefont {G.}~\bibnamefont
  {Baur}}, \bibinfo {author} {\bibfnamefont {G.}~\bibnamefont {Boero}},
  \bibinfo {author} {\bibfnamefont {A.}~\bibnamefont {Brauksiepe}}, \bibinfo
  {author} {\bibfnamefont {A.}~\bibnamefont {Buzzo}}, \bibinfo {author}
  {\bibfnamefont {W.}~\bibnamefont {Eyrich}}, \bibinfo {author} {\bibfnamefont
  {R.}~\bibnamefont {Geyer}}, \bibinfo {author} {\bibfnamefont
  {D.}~\bibnamefont {Grzonka}}, \bibinfo {author} {\bibfnamefont
  {J.}~\bibnamefont {Hauffe}}, \bibinfo {author} {\bibfnamefont
  {K.}~\bibnamefont {Kilian}}, \bibinfo {author} {\bibfnamefont
  {M.}~\bibnamefont {LoVetere}}, \bibinfo {author} {\bibfnamefont
  {M.}~\bibnamefont {Macri}}, \bibinfo {author} {\bibfnamefont
  {M.}~\bibnamefont {Moosburger}}, \bibinfo {author} {\bibfnamefont
  {R.}~\bibnamefont {Nellen}}, \bibinfo {author} {\bibfnamefont
  {W.}~\bibnamefont {Oelert}}, \bibinfo {author} {\bibfnamefont
  {S.}~\bibnamefont {Passaggio}}, \bibinfo {author} {\bibfnamefont
  {A.}~\bibnamefont {Pozzo}}, \bibinfo {author} {\bibfnamefont
  {K.}~\bibnamefont {R{\"o}hrich}}, \bibinfo {author} {\bibfnamefont
  {K.}~\bibnamefont {Sachs}}, \bibinfo {author} {\bibfnamefont
  {G.}~\bibnamefont {Schepers}}, \bibinfo {author} {\bibfnamefont
  {T.}~\bibnamefont {Sefzick}}, \bibinfo {author} {\bibfnamefont
  {R.}~\bibnamefont {Simon}}, \bibinfo {author} {\bibfnamefont
  {R.}~\bibnamefont {Stratmann}}, \bibinfo {author} {\bibfnamefont
  {F.}~\bibnamefont {Stinzing}},\ and\ \bibinfo {author} {\bibfnamefont
  {M.}~\bibnamefont {Wolke}},\ }\bibfield  {title} {\bibinfo {title}
  {Production of antihydrogen},\ }\href@noop {} {\bibfield  {journal} {\bibinfo
   {journal} {Phys.\,Lett.\,B}\ }\textbf {\bibinfo {volume} {368}},\ \bibinfo
  {pages} {251} (\bibinfo {year} {1996})}\BibitemShut {NoStop}%
\bibitem [{\citenamefont {Blanford}\ \emph {et~al.}(1998)\citenamefont
  {Blanford}, \citenamefont {Christian}, \citenamefont {Gollwitzer},
  \citenamefont {Mandelkern}, \citenamefont {Munger}, \citenamefont {Schultz},\
  and\ \citenamefont {Zioulas}}]{anti:blan98}%
  \BibitemOpen
  \bibfield  {author} {\bibinfo {author} {\bibfnamefont {G.}~\bibnamefont
  {Blanford}}, \bibinfo {author} {\bibfnamefont {D.~C.}\ \bibnamefont
  {Christian}}, \bibinfo {author} {\bibfnamefont {K.}~\bibnamefont
  {Gollwitzer}}, \bibinfo {author} {\bibfnamefont {M.}~\bibnamefont
  {Mandelkern}}, \bibinfo {author} {\bibfnamefont {C.~T.}\ \bibnamefont
  {Munger}}, \bibinfo {author} {\bibfnamefont {J.}~\bibnamefont {Schultz}},\
  and\ \bibinfo {author} {\bibfnamefont {G.}~\bibnamefont {Zioulas}},\
  }\bibfield  {title} {\bibinfo {title} {Observation of atomic antihydrogen},\
  }\href@noop {} {\bibfield  {journal} {\bibinfo  {journal}
  {Phys.\,Rev.\,Lett.}\ }\textbf {\bibinfo {volume} {80}},\ \bibinfo {pages}
  {3037} (\bibinfo {year} {1998})}\BibitemShut {NoStop}%
\bibitem [{\citenamefont {Amoretti}\ \emph {et~al.}(2002)\citenamefont
  {Amoretti}, \citenamefont {Amsler}, \citenamefont {Bonomi}, \citenamefont
  {Bouchta}, \citenamefont {Bowe}, \citenamefont {Carraro}, \citenamefont
  {Cesar}, \citenamefont {Charlton}, \citenamefont {Collier},\ and\
  \citenamefont {et~al.}}]{anti:amor02}%
  \BibitemOpen
  \bibfield  {author} {\bibinfo {author} {\bibfnamefont {M.}~\bibnamefont
  {Amoretti}}, \bibinfo {author} {\bibfnamefont {C.}~\bibnamefont {Amsler}},
  \bibinfo {author} {\bibfnamefont {G.}~\bibnamefont {Bonomi}}, \bibinfo
  {author} {\bibfnamefont {A.}~\bibnamefont {Bouchta}}, \bibinfo {author}
  {\bibfnamefont {P.}~\bibnamefont {Bowe}}, \bibinfo {author} {\bibfnamefont
  {C.}~\bibnamefont {Carraro}}, \bibinfo {author} {\bibfnamefont {C.~L.}\
  \bibnamefont {Cesar}}, \bibinfo {author} {\bibfnamefont {M.}~\bibnamefont
  {Charlton}}, \bibinfo {author} {\bibfnamefont {M.~J.~T.}\ \bibnamefont
  {Collier}},\ and\ \bibinfo {author} {\bibfnamefont {M.~D.}\ \bibnamefont
  {et~al.}},\ }\bibfield  {title} {\bibinfo {title} {Production and detection
  of cold antihydrogen atoms},\ }\href@noop {} {\bibfield  {journal} {\bibinfo
  {journal} {Nature}\ }\textbf {\bibinfo {volume} {419}},\ \bibinfo {pages}
  {456} (\bibinfo {year} {2002})}\BibitemShut {NoStop}%
\bibitem [{\citenamefont {Gabrielse}\ \emph
  {et~al.}(2002{\natexlab{a}})\citenamefont {Gabrielse}, \citenamefont
  {Bowden}, \citenamefont {Oxley}, \citenamefont {Speck}, \citenamefont
  {Storry}, \citenamefont {Tan}, \citenamefont {Wessels}, \citenamefont
  {Grzonka}, \citenamefont {Oelert}, \citenamefont {Schepers}, \citenamefont
  {Sefzick}, \citenamefont {Walz}, \citenamefont {Pittner}, \citenamefont
  {H\"ansch},\ and\ \citenamefont {Hessels}}]{anti:gabr02}%
  \BibitemOpen
  \bibfield  {author} {\bibinfo {author} {\bibfnamefont {G.}~\bibnamefont
  {Gabrielse}}, \bibinfo {author} {\bibfnamefont {N.~S.}\ \bibnamefont
  {Bowden}}, \bibinfo {author} {\bibfnamefont {P.}~\bibnamefont {Oxley}},
  \bibinfo {author} {\bibfnamefont {A.}~\bibnamefont {Speck}}, \bibinfo
  {author} {\bibfnamefont {C.~H.}\ \bibnamefont {Storry}}, \bibinfo {author}
  {\bibfnamefont {J.~N.}\ \bibnamefont {Tan}}, \bibinfo {author} {\bibfnamefont
  {M.}~\bibnamefont {Wessels}}, \bibinfo {author} {\bibfnamefont
  {D.}~\bibnamefont {Grzonka}}, \bibinfo {author} {\bibfnamefont
  {W.}~\bibnamefont {Oelert}}, \bibinfo {author} {\bibfnamefont
  {G.}~\bibnamefont {Schepers}}, \bibinfo {author} {\bibfnamefont
  {T.}~\bibnamefont {Sefzick}}, \bibinfo {author} {\bibfnamefont
  {J.}~\bibnamefont {Walz}}, \bibinfo {author} {\bibfnamefont {H.}~\bibnamefont
  {Pittner}}, \bibinfo {author} {\bibfnamefont {T.~W.}\ \bibnamefont
  {H\"ansch}},\ and\ \bibinfo {author} {\bibfnamefont {E.~A.}\ \bibnamefont
  {Hessels}},\ }\bibfield  {title} {\bibinfo {title} {Background-free
  observation of cold antihydrogen with field-ionization analysis of its
  states},\ }\href@noop {} {\bibfield  {journal} {\bibinfo  {journal}
  {Phys.\,Rev.\,Lett.}\ }\textbf {\bibinfo {volume} {89}},\ \bibinfo {pages}
  {213401} (\bibinfo {year} {2002}{\natexlab{a}})}\BibitemShut {NoStop}%
\bibitem [{\citenamefont {Gabrielse}\ \emph
  {et~al.}(2002{\natexlab{b}})\citenamefont {Gabrielse} \emph
  {et~al.}}]{anti:gabr02a}%
  \BibitemOpen
  \bibfield  {author} {\bibinfo {author} {\bibfnamefont {G.}~\bibnamefont
  {Gabrielse}} \emph {et~al.},\ }\bibfield  {title} {\bibinfo {title} {Driven
  production of cold antihydrogen and the first measured distribution of
  antihydrogen states},\ }\href@noop {} {\bibfield  {journal} {\bibinfo
  {journal} {Phys.\,Rev.\,Lett.}\ }\textbf {\bibinfo {volume} {89}},\ \bibinfo
  {pages} {233401} (\bibinfo {year} {2002}{\natexlab{b}})}\BibitemShut
  {NoStop}%
\bibitem [{\citenamefont {Bertsche}\ \emph {et~al.}(2015)\citenamefont
  {Bertsche}, \citenamefont {Butler}, \citenamefont {Charlton},\ and\
  \citenamefont {Madsen}}]{anti:bert15}%
  \BibitemOpen
  \bibfield  {author} {\bibinfo {author} {\bibfnamefont {W.~A.}\ \bibnamefont
  {Bertsche}}, \bibinfo {author} {\bibfnamefont {E.}~\bibnamefont {Butler}},
  \bibinfo {author} {\bibfnamefont {M.}~\bibnamefont {Charlton}},\ and\
  \bibinfo {author} {\bibfnamefont {N.}~\bibnamefont {Madsen}},\ }\bibfield
  {title} {\bibinfo {title} {Physics with antihydrogen},\ }\href@noop {}
  {\bibfield  {journal} {\bibinfo  {journal} {J.\,Phys.\,B}\ }\textbf {\bibinfo
  {volume} {48}},\ \bibinfo {pages} {232001} (\bibinfo {year}
  {2015})}\BibitemShut {NoStop}%
\bibitem [{\citenamefont {Ahmadi}\ \emph {et~al.}(2018)\citenamefont {Ahmadi},
  \citenamefont {Alves}, \citenamefont {Baker}, \citenamefont {Bertsche},
  \citenamefont {Capra}, \citenamefont {Carruth}, \citenamefont {Cesar},
  \citenamefont {Charlton}, \citenamefont {Cohen}, \citenamefont {Collister},
  \citenamefont {Eriksson}, \citenamefont {Evans}, \citenamefont {Evetts},
  \citenamefont {Fajans}, \citenamefont {Friesen}, \citenamefont {Fujiwara},
  \citenamefont {Gill}, \citenamefont {Hangst}, \citenamefont {Hardy},
  \citenamefont {Hayden}, \citenamefont {Isaac}, \citenamefont {Johnson},
  \citenamefont {Jones}, \citenamefont {Jones}, \citenamefont {Jonsell},
  \citenamefont {Khramov}, \citenamefont {Kurchaninov}, \citenamefont {Madsen},
  \citenamefont {Maxwell}, \citenamefont {McKenna}, \citenamefont {Menary},
  \citenamefont {Momose}, \citenamefont {Munich}, \citenamefont {Olchanski},
  \citenamefont {Olin}, \citenamefont {Pusa}, \citenamefont {Rasmussen},
  \citenamefont {Robicheaux}, \citenamefont {Sacramento}, \citenamefont
  {Sameed}, \citenamefont {Sarid}, \citenamefont {Silveira}, \citenamefont
  {Stutter}, \citenamefont {So}, \citenamefont {Tharp}, \citenamefont
  {Thompson}, \citenamefont {van~der Werf},\ and\ \citenamefont
  {Wurtele}}]{anti:ahma18}%
  \BibitemOpen
  \bibfield  {author} {\bibinfo {author} {\bibfnamefont {M.}~\bibnamefont
  {Ahmadi}}, \bibinfo {author} {\bibfnamefont {B.~X.~R.}\ \bibnamefont
  {Alves}}, \bibinfo {author} {\bibfnamefont {C.~J.}\ \bibnamefont {Baker}},
  \bibinfo {author} {\bibfnamefont {W.}~\bibnamefont {Bertsche}}, \bibinfo
  {author} {\bibfnamefont {A.}~\bibnamefont {Capra}}, \bibinfo {author}
  {\bibfnamefont {C.}~\bibnamefont {Carruth}}, \bibinfo {author} {\bibfnamefont
  {C.~L.}\ \bibnamefont {Cesar}}, \bibinfo {author} {\bibfnamefont
  {M.}~\bibnamefont {Charlton}}, \bibinfo {author} {\bibfnamefont
  {S.}~\bibnamefont {Cohen}}, \bibinfo {author} {\bibfnamefont
  {R.}~\bibnamefont {Collister}}, \bibinfo {author} {\bibfnamefont
  {S.}~\bibnamefont {Eriksson}}, \bibinfo {author} {\bibfnamefont
  {A.}~\bibnamefont {Evans}}, \bibinfo {author} {\bibfnamefont
  {N.}~\bibnamefont {Evetts}}, \bibinfo {author} {\bibfnamefont
  {J.}~\bibnamefont {Fajans}}, \bibinfo {author} {\bibfnamefont
  {T.}~\bibnamefont {Friesen}}, \bibinfo {author} {\bibfnamefont {M.~C.}\
  \bibnamefont {Fujiwara}}, \bibinfo {author} {\bibfnamefont {D.~R.}\
  \bibnamefont {Gill}}, \bibinfo {author} {\bibfnamefont {J.~S.}\ \bibnamefont
  {Hangst}}, \bibinfo {author} {\bibfnamefont {W.~N.}\ \bibnamefont {Hardy}},
  \bibinfo {author} {\bibfnamefont {M.~E.}\ \bibnamefont {Hayden}}, \bibinfo
  {author} {\bibfnamefont {C.~A.}\ \bibnamefont {Isaac}}, \bibinfo {author}
  {\bibfnamefont {M.~A.}\ \bibnamefont {Johnson}}, \bibinfo {author}
  {\bibfnamefont {J.~M.}\ \bibnamefont {Jones}}, \bibinfo {author}
  {\bibfnamefont {S.~A.}\ \bibnamefont {Jones}}, \bibinfo {author}
  {\bibfnamefont {S.}~\bibnamefont {Jonsell}}, \bibinfo {author} {\bibfnamefont
  {P.}~\bibnamefont {Khramov}, \bibfnamefont {A.and~Knapp}}, \bibinfo {author}
  {\bibfnamefont {L.}~\bibnamefont {Kurchaninov}}, \bibinfo {author}
  {\bibfnamefont {N.}~\bibnamefont {Madsen}}, \bibinfo {author} {\bibfnamefont
  {D.}~\bibnamefont {Maxwell}}, \bibinfo {author} {\bibfnamefont {J.~T.~K.}\
  \bibnamefont {McKenna}}, \bibinfo {author} {\bibfnamefont {S.}~\bibnamefont
  {Menary}}, \bibinfo {author} {\bibfnamefont {T.}~\bibnamefont {Momose}},
  \bibinfo {author} {\bibfnamefont {J.~J.}\ \bibnamefont {Munich}}, \bibinfo
  {author} {\bibfnamefont {K.}~\bibnamefont {Olchanski}}, \bibinfo {author}
  {\bibfnamefont {A.}~\bibnamefont {Olin}}, \bibinfo {author} {\bibfnamefont
  {P.}~\bibnamefont {Pusa}}, \bibinfo {author} {\bibfnamefont {C.~{\O}.}\
  \bibnamefont {Rasmussen}}, \bibinfo {author} {\bibfnamefont {F.}~\bibnamefont
  {Robicheaux}}, \bibinfo {author} {\bibfnamefont {R.~L.}\ \bibnamefont
  {Sacramento}}, \bibinfo {author} {\bibfnamefont {M.}~\bibnamefont {Sameed}},
  \bibinfo {author} {\bibfnamefont {E.}~\bibnamefont {Sarid}}, \bibinfo
  {author} {\bibfnamefont {D.~M.}\ \bibnamefont {Silveira}}, \bibinfo {author}
  {\bibfnamefont {G.}~\bibnamefont {Stutter}}, \bibinfo {author} {\bibfnamefont
  {C.}~\bibnamefont {So}}, \bibinfo {author} {\bibfnamefont {T.~D.}\
  \bibnamefont {Tharp}}, \bibinfo {author} {\bibfnamefont {R.~I.}\ \bibnamefont
  {Thompson}}, \bibinfo {author} {\bibfnamefont {D.~P.}\ \bibnamefont {van~der
  Werf}},\ and\ \bibinfo {author} {\bibfnamefont {J.~S.}\ \bibnamefont
  {Wurtele}},\ }\bibfield  {title} {\bibinfo {title} {Characterization of the
  1s-2s transition in antihydrogen},\ }\href@noop {} {\bibfield  {journal}
  {\bibinfo  {journal} {Nature}\ }\textbf {\bibinfo {volume} {557}},\ \bibinfo
  {pages} {71} (\bibinfo {year} {2018})}\BibitemShut {NoStop}%
\bibitem [{\citenamefont {Baker}\ \emph {et~al.}(2025)\citenamefont {Baker},
  \citenamefont {Bertsche}, \citenamefont {Capra}, \citenamefont {Carruth},
  \citenamefont {Cesar}, \citenamefont {Charlton}, \citenamefont {Christensen},
  \citenamefont {Collister}, \citenamefont {Cridland~Mathad}, \citenamefont
  {Eriksson}, \citenamefont {Evans}, \citenamefont {Evetts}, \citenamefont
  {Fajans}, \citenamefont {Friesen}, \citenamefont {Fujiwara}, \citenamefont
  {Gill}, \citenamefont {Grandemange}, \citenamefont {Granum}, \citenamefont
  {Hangst}, \citenamefont {Hardy}, \citenamefont {Hayden}, \citenamefont
  {Hodgkinson}, \citenamefont {Hunter}, \citenamefont {Isaac}, \citenamefont
  {Johnson}, \citenamefont {Jones}, \citenamefont {Jones}, \citenamefont
  {Jonsell}, \citenamefont {Khramov}, \citenamefont {Kurchaninov},
  \citenamefont {Madsen}, \citenamefont {Maxwell}, \citenamefont {McKenna},
  \citenamefont {Menary}, \citenamefont {Momose}, \citenamefont {Mullan},
  \citenamefont {Munich}, \citenamefont {Olchanski}, \citenamefont {Olin}, ,
  \citenamefont {Peszka}, \citenamefont {Powell}, \citenamefont {Pusa},
  \citenamefont {Rasmussen}, \citenamefont {Robicheaux}, \citenamefont
  {Sacramento}, \citenamefont {Sameed}, \citenamefont {Sarid}, \citenamefont
  {Silveira}, \citenamefont {So}, \citenamefont {Stutter}, \citenamefont
  {Tharp}, \citenamefont {Thompson}, \citenamefont {van~der Werf},
  \citenamefont {Wurtele},\ and\ \citenamefont {Shore}}]{anti:bake25}%
  \BibitemOpen
  \bibfield  {author} {\bibinfo {author} {\bibfnamefont {C.~J.}\ \bibnamefont
  {Baker}}, \bibinfo {author} {\bibfnamefont {W.}~\bibnamefont {Bertsche}},
  \bibinfo {author} {\bibfnamefont {A.}~\bibnamefont {Capra}}, \bibinfo
  {author} {\bibfnamefont {C.}~\bibnamefont {Carruth}}, \bibinfo {author}
  {\bibfnamefont {C.~L.}\ \bibnamefont {Cesar}}, \bibinfo {author}
  {\bibfnamefont {M.}~\bibnamefont {Charlton}}, \bibinfo {author}
  {\bibfnamefont {A.}~\bibnamefont {Christensen}}, \bibinfo {author}
  {\bibfnamefont {R.}~\bibnamefont {Collister}}, \bibinfo {author}
  {\bibfnamefont {A.}~\bibnamefont {Cridland~Mathad}}, \bibinfo {author}
  {\bibfnamefont {S.}~\bibnamefont {Eriksson}}, \bibinfo {author}
  {\bibfnamefont {A.}~\bibnamefont {Evans}}, \bibinfo {author} {\bibfnamefont
  {N.}~\bibnamefont {Evetts}}, \bibinfo {author} {\bibfnamefont
  {J.}~\bibnamefont {Fajans}}, \bibinfo {author} {\bibfnamefont
  {T.}~\bibnamefont {Friesen}}, \bibinfo {author} {\bibfnamefont {M.~C.}\
  \bibnamefont {Fujiwara}}, \bibinfo {author} {\bibfnamefont {D.~R.}\
  \bibnamefont {Gill}}, \bibinfo {author} {\bibfnamefont {P.}~\bibnamefont
  {Grandemange}}, \bibinfo {author} {\bibfnamefont {P.}~\bibnamefont {Granum}},
  \bibinfo {author} {\bibfnamefont {J.~S.}\ \bibnamefont {Hangst}}, \bibinfo
  {author} {\bibfnamefont {W.~N.}\ \bibnamefont {Hardy}}, \bibinfo {author}
  {\bibfnamefont {M.~E.}\ \bibnamefont {Hayden}}, \bibinfo {author}
  {\bibfnamefont {D.}~\bibnamefont {Hodgkinson}}, \bibinfo {author}
  {\bibfnamefont {E.}~\bibnamefont {Hunter}}, \bibinfo {author} {\bibfnamefont
  {C.~A.}\ \bibnamefont {Isaac}}, \bibinfo {author} {\bibfnamefont {M.~A.}\
  \bibnamefont {Johnson}}, \bibinfo {author} {\bibfnamefont {J.~M.}\
  \bibnamefont {Jones}}, \bibinfo {author} {\bibfnamefont {S.~A.}\ \bibnamefont
  {Jones}}, \bibinfo {author} {\bibfnamefont {S.}~\bibnamefont {Jonsell}},
  \bibinfo {author} {\bibfnamefont {A.}~\bibnamefont {Khramov}}, \bibinfo
  {author} {\bibfnamefont {L.}~\bibnamefont {Kurchaninov}}, \bibinfo {author}
  {\bibfnamefont {N.}~\bibnamefont {Madsen}}, \bibinfo {author} {\bibfnamefont
  {D.}~\bibnamefont {Maxwell}}, \bibinfo {author} {\bibfnamefont {J.~T.~K.}\
  \bibnamefont {McKenna}}, \bibinfo {author} {\bibfnamefont {S.}~\bibnamefont
  {Menary}}, \bibinfo {author} {\bibfnamefont {T.}~\bibnamefont {Momose}},
  \bibinfo {author} {\bibfnamefont {P.~S.}\ \bibnamefont {Mullan}}, \bibinfo
  {author} {\bibfnamefont {J.~J.}\ \bibnamefont {Munich}}, \bibinfo {author}
  {\bibfnamefont {K.}~\bibnamefont {Olchanski}}, \bibinfo {author}
  {\bibfnamefont {A.}~\bibnamefont {Olin}}, , \bibinfo {author} {\bibfnamefont
  {J.}~\bibnamefont {Peszka}}, \bibinfo {author} {\bibfnamefont
  {A.}~\bibnamefont {Powell}}, \bibinfo {author} {\bibfnamefont
  {P.}~\bibnamefont {Pusa}}, \bibinfo {author} {\bibfnamefont {C.~{\O}.}\
  \bibnamefont {Rasmussen}}, \bibinfo {author} {\bibfnamefont {F.}~\bibnamefont
  {Robicheaux}}, \bibinfo {author} {\bibfnamefont {R.~L.}\ \bibnamefont
  {Sacramento}}, \bibinfo {author} {\bibfnamefont {M.}~\bibnamefont {Sameed}},
  \bibinfo {author} {\bibfnamefont {E.}~\bibnamefont {Sarid}}, \bibinfo
  {author} {\bibfnamefont {D.~M.}\ \bibnamefont {Silveira}}, \bibinfo {author}
  {\bibfnamefont {C.}~\bibnamefont {So}}, \bibinfo {author} {\bibfnamefont
  {G.}~\bibnamefont {Stutter}}, \bibinfo {author} {\bibfnamefont {T.~D.}\
  \bibnamefont {Tharp}}, \bibinfo {author} {\bibfnamefont {R.~I.}\ \bibnamefont
  {Thompson}}, \bibinfo {author} {\bibfnamefont {D.~P.}\ \bibnamefont {van~der
  Werf}}, \bibinfo {author} {\bibfnamefont {J.~S.}\ \bibnamefont {Wurtele}},\
  and\ \bibinfo {author} {\bibfnamefont {G.~M.}\ \bibnamefont {Shore}},\
  }\bibfield  {title} {\bibinfo {title} {Precision spectroscopy of the
  hyperfine components of the 1s–2s transition in antihydrogen},\ }\href@noop
  {} {\bibfield  {journal} {\bibinfo  {journal} {Nat.\,Phys.}\ }\textbf
  {\bibinfo {volume} {21}},\ \bibinfo {pages} {201} (\bibinfo {year}
  {2025})}\BibitemShut {NoStop}%
\bibitem [{\citenamefont {Anderson}\ \emph {et~al.}(2023)\citenamefont
  {Anderson}, \citenamefont {Baker}, \citenamefont {Bertsche}, \citenamefont
  {Bhatt}, \citenamefont {Bonomi}, \citenamefont {Capra}, \citenamefont
  {Carli}, \citenamefont {Cesar}, \citenamefont {Charlton}, \citenamefont
  {Christensen}, \citenamefont {Collister}, \citenamefont {Cridland~Mathad},
  \citenamefont {Duque~Quiceno}, \citenamefont {Eriksson}, \citenamefont
  {Evans}, \citenamefont {Evetts}, \citenamefont {Fabbri}, \citenamefont
  {Fajans}, \citenamefont {Ferwerda}, \citenamefont {Friesen}, , \citenamefont
  {Fujiwara}, \citenamefont {Gill}, \citenamefont {Golino}, \citenamefont
  {Gomes~Gonçalves}, \citenamefont {Grandemange}, \citenamefont {Granum},
  \citenamefont {Hangst}, \citenamefont {Hayden}, \citenamefont {Hodgkinson},
  \citenamefont {Hunter}, \citenamefont {Isaac}, \citenamefont {Jimenez},
  \citenamefont {Johnson}, \citenamefont {Jones}, \citenamefont {Jones},
  \citenamefont {Jonsell}, \citenamefont {Khramov}, \citenamefont {Madsen},
  \citenamefont {Martin}, \citenamefont {Massacret}, \citenamefont {Maxwell},
  \citenamefont {McKenna}, \citenamefont {Menary}, \citenamefont {Momose},
  \citenamefont {Mostamand}, \citenamefont {Mullan}, \citenamefont {Nauta},
  \citenamefont {Olchanski}, \citenamefont {Oliveira}, \citenamefont {Peszka},
  \citenamefont {Powell}, \citenamefont {Rasmussen}, \citenamefont
  {Robicheaux}, \citenamefont {Sacramento}, \citenamefont {Sameed},
  \citenamefont {Sarid}, \citenamefont {Schoonwater}, \citenamefont {Silveira},
  \citenamefont {Singh}, \citenamefont {Smith}, \citenamefont {So},
  \citenamefont {Stracka}, \citenamefont {Stutter}, \citenamefont {Tharp},
  \citenamefont {Thompson}, \citenamefont {Thompson}, \citenamefont
  {Thorpe-Woods}, \citenamefont {Torkzaban}, \citenamefont {Urioni},
  \citenamefont {Woosaree},\ and\ \citenamefont {Wurtele}}]{anti:ande23}%
  \BibitemOpen
  \bibfield  {author} {\bibinfo {author} {\bibfnamefont {E.~K.}\ \bibnamefont
  {Anderson}}, \bibinfo {author} {\bibfnamefont {C.~J.}\ \bibnamefont {Baker}},
  \bibinfo {author} {\bibfnamefont {W.}~\bibnamefont {Bertsche}}, \bibinfo
  {author} {\bibfnamefont {N.~M.}\ \bibnamefont {Bhatt}}, \bibinfo {author}
  {\bibfnamefont {G.}~\bibnamefont {Bonomi}}, \bibinfo {author} {\bibfnamefont
  {A.}~\bibnamefont {Capra}}, \bibinfo {author} {\bibfnamefont
  {I.}~\bibnamefont {Carli}}, \bibinfo {author} {\bibfnamefont {C.~L.}\
  \bibnamefont {Cesar}}, \bibinfo {author} {\bibfnamefont {M.}~\bibnamefont
  {Charlton}}, \bibinfo {author} {\bibfnamefont {A.}~\bibnamefont
  {Christensen}}, \bibinfo {author} {\bibfnamefont {R.}~\bibnamefont
  {Collister}}, \bibinfo {author} {\bibfnamefont {A.}~\bibnamefont
  {Cridland~Mathad}}, \bibinfo {author} {\bibfnamefont {D.}~\bibnamefont
  {Duque~Quiceno}}, \bibinfo {author} {\bibfnamefont {S.}~\bibnamefont
  {Eriksson}}, \bibinfo {author} {\bibfnamefont {A.}~\bibnamefont {Evans}},
  \bibinfo {author} {\bibfnamefont {N.}~\bibnamefont {Evetts}}, \bibinfo
  {author} {\bibfnamefont {S.}~\bibnamefont {Fabbri}}, \bibinfo {author}
  {\bibfnamefont {J.}~\bibnamefont {Fajans}}, \bibinfo {author} {\bibfnamefont
  {A.}~\bibnamefont {Ferwerda}}, \bibinfo {author} {\bibfnamefont
  {T.}~\bibnamefont {Friesen}}, , \bibinfo {author} {\bibfnamefont {M.~C.}\
  \bibnamefont {Fujiwara}}, \bibinfo {author} {\bibfnamefont {D.~R.}\
  \bibnamefont {Gill}}, \bibinfo {author} {\bibfnamefont {L.~M.}\ \bibnamefont
  {Golino}}, \bibinfo {author} {\bibfnamefont {M.~B.}\ \bibnamefont
  {Gomes~Gonçalves}}, \bibinfo {author} {\bibfnamefont {P.}~\bibnamefont
  {Grandemange}}, \bibinfo {author} {\bibfnamefont {P.}~\bibnamefont {Granum}},
  \bibinfo {author} {\bibfnamefont {J.~S.}\ \bibnamefont {Hangst}}, \bibinfo
  {author} {\bibfnamefont {M.~E.}\ \bibnamefont {Hayden}}, \bibinfo {author}
  {\bibfnamefont {D.}~\bibnamefont {Hodgkinson}}, \bibinfo {author}
  {\bibfnamefont {E.~D.}\ \bibnamefont {Hunter}}, \bibinfo {author}
  {\bibfnamefont {C.~A.}\ \bibnamefont {Isaac}}, \bibinfo {author}
  {\bibfnamefont {A.~J.~U.}\ \bibnamefont {Jimenez}}, \bibinfo {author}
  {\bibfnamefont {M.~A.}\ \bibnamefont {Johnson}}, \bibinfo {author}
  {\bibfnamefont {J.~M.}\ \bibnamefont {Jones}}, \bibinfo {author}
  {\bibfnamefont {S.~A.}\ \bibnamefont {Jones}}, \bibinfo {author}
  {\bibfnamefont {S.}~\bibnamefont {Jonsell}}, \bibinfo {author} {\bibfnamefont
  {A.}~\bibnamefont {Khramov}}, \bibinfo {author} {\bibfnamefont
  {N.}~\bibnamefont {Madsen}}, \bibinfo {author} {\bibfnamefont
  {L.}~\bibnamefont {Martin}}, \bibinfo {author} {\bibfnamefont
  {N.}~\bibnamefont {Massacret}}, \bibinfo {author} {\bibfnamefont
  {D.}~\bibnamefont {Maxwell}}, \bibinfo {author} {\bibfnamefont {J.~T.~K.}\
  \bibnamefont {McKenna}}, \bibinfo {author} {\bibfnamefont {S.}~\bibnamefont
  {Menary}}, \bibinfo {author} {\bibfnamefont {T.}~\bibnamefont {Momose}},
  \bibinfo {author} {\bibfnamefont {M.}~\bibnamefont {Mostamand}}, \bibinfo
  {author} {\bibfnamefont {P.~S.}\ \bibnamefont {Mullan}}, \bibinfo {author}
  {\bibfnamefont {J.}~\bibnamefont {Nauta}}, \bibinfo {author} {\bibfnamefont
  {K.}~\bibnamefont {Olchanski}}, \bibinfo {author} {\bibfnamefont {A.~N.}\
  \bibnamefont {Oliveira}}, \bibinfo {author} {\bibfnamefont {J.}~\bibnamefont
  {Peszka}}, \bibinfo {author} {\bibfnamefont {A.}~\bibnamefont {Powell}},
  \bibinfo {author} {\bibfnamefont {C.~{\O}.}\ \bibnamefont {Rasmussen}},
  \bibinfo {author} {\bibfnamefont {F.}~\bibnamefont {Robicheaux}}, \bibinfo
  {author} {\bibfnamefont {R.~L.}\ \bibnamefont {Sacramento}}, \bibinfo
  {author} {\bibfnamefont {M.}~\bibnamefont {Sameed}}, \bibinfo {author}
  {\bibfnamefont {E.}~\bibnamefont {Sarid}}, \bibinfo {author} {\bibfnamefont
  {J.}~\bibnamefont {Schoonwater}}, \bibinfo {author} {\bibfnamefont {D.~M.}\
  \bibnamefont {Silveira}}, \bibinfo {author} {\bibfnamefont {J.}~\bibnamefont
  {Singh}}, \bibinfo {author} {\bibfnamefont {G.}~\bibnamefont {Smith}},
  \bibinfo {author} {\bibfnamefont {C.}~\bibnamefont {So}}, \bibinfo {author}
  {\bibfnamefont {S.}~\bibnamefont {Stracka}}, \bibinfo {author} {\bibfnamefont
  {G.}~\bibnamefont {Stutter}}, \bibinfo {author} {\bibfnamefont {T.~D.}\
  \bibnamefont {Tharp}}, \bibinfo {author} {\bibfnamefont {K.~A.}\ \bibnamefont
  {Thompson}}, \bibinfo {author} {\bibfnamefont {R.~I.}\ \bibnamefont
  {Thompson}}, \bibinfo {author} {\bibfnamefont {E.}~\bibnamefont
  {Thorpe-Woods}}, \bibinfo {author} {\bibfnamefont {C.}~\bibnamefont
  {Torkzaban}}, \bibinfo {author} {\bibfnamefont {M.}~\bibnamefont {Urioni}},
  \bibinfo {author} {\bibfnamefont {P.}~\bibnamefont {Woosaree}},\ and\
  \bibinfo {author} {\bibfnamefont {J.~S.}\ \bibnamefont {Wurtele}},\
  }\bibfield  {title} {\bibinfo {title} {Observation of the effect of gravity
  on the motion of antimatter},\ }\href@noop {} {\bibfield  {journal} {\bibinfo
   {journal} {Nature}\ }\textbf {\bibinfo {volume} {621}},\ \bibinfo {pages}
  {716} (\bibinfo {year} {2023})}\BibitemShut {NoStop}%
\bibitem [{\citenamefont {Ko{\l o}s}\ \emph {et~al.}(1975)\citenamefont {Ko{\l
  o}s}, \citenamefont {Morgan}, \citenamefont {Schrader},\ and\ \citenamefont
  {Wolniewicz}}]{anti:kolo75}%
  \BibitemOpen
  \bibfield  {author} {\bibinfo {author} {\bibfnamefont {W.}~\bibnamefont
  {Ko{\l o}s}}, \bibinfo {author} {\bibfnamefont {D.~L.}\ \bibnamefont
  {Morgan}}, \bibinfo {author} {\bibfnamefont {D.~M.}\ \bibnamefont
  {Schrader}},\ and\ \bibinfo {author} {\bibfnamefont {L.}~\bibnamefont
  {Wolniewicz}},\ }\bibfield  {title} {\bibinfo {title} {Hydrogen-antihydrogen
  interactions},\ }\href@noop {} {\bibfield  {journal} {\bibinfo  {journal}
  {Phys.\,Rev.\,A}\ }\textbf {\bibinfo {volume} {11}},\ \bibinfo {pages} {1792}
  (\bibinfo {year} {1975})}\BibitemShut {NoStop}%
\bibitem [{\citenamefont {Morgan}\ and\ \citenamefont
  {Hughes}(1973)}]{anti:morg73}%
  \BibitemOpen
  \bibfield  {author} {\bibinfo {author} {\bibfnamefont {D.~L.}\ \bibnamefont
  {Morgan}}\ and\ \bibinfo {author} {\bibfnamefont {V.~W.}\ \bibnamefont
  {Hughes}},\ }\bibfield  {title} {\bibinfo {title} {Atom-antiatom
  interactions},\ }\href@noop {} {\bibfield  {journal} {\bibinfo  {journal}
  {Phys.\,Rev.\,A}\ }\textbf {\bibinfo {volume} {7}},\ \bibinfo {pages} {1811}
  (\bibinfo {year} {1973})}\BibitemShut {NoStop}%
\bibitem [{\citenamefont {Fermi}\ and\ \citenamefont
  {Teller}(1947)}]{anti:ferm47}%
  \BibitemOpen
  \bibfield  {author} {\bibinfo {author} {\bibfnamefont {E.}~\bibnamefont
  {Fermi}}\ and\ \bibinfo {author} {\bibfnamefont {E.}~\bibnamefont {Teller}},\
  }\bibfield  {title} {\bibinfo {title} {The capture of negative mesotrons in
  matter},\ }\href@noop {} {\bibfield  {journal} {\bibinfo  {journal}
  {Phys.\,Rev.}\ }\textbf {\bibinfo {volume} {72}},\ \bibinfo {pages} {399}
  (\bibinfo {year} {1947})}\BibitemShut {NoStop}%
\bibitem [{\citenamefont {Gridnev}\ and\ \citenamefont
  {Greiner}(2005)}]{anti:grid05}%
  \BibitemOpen
  \bibfield  {author} {\bibinfo {author} {\bibfnamefont {D.~K.}\ \bibnamefont
  {Gridnev}}\ and\ \bibinfo {author} {\bibfnamefont {C.}~\bibnamefont
  {Greiner}},\ }\bibfield  {title} {\bibinfo {title} {Proof that the
  hydrogen-antihydrogen molecule is unstable},\ }\href@noop {} {\bibfield
  {journal} {\bibinfo  {journal} {Phys.\,Rev.\,Lett.}\ }\textbf {\bibinfo
  {volume} {94}},\ \bibinfo {pages} {223402} (\bibinfo {year}
  {2005})}\BibitemShut {NoStop}%
\bibitem [{\citenamefont {Saenz}(2006)}]{anti:saen06}%
  \BibitemOpen
  \bibfield  {author} {\bibinfo {author} {\bibfnamefont {A.}~\bibnamefont
  {Saenz}},\ }\bibfield  {title} {\bibinfo {title} {Density-matrix study of the
  hydrogen-antihydrogen molecule},\ }\href@noop {} {\bibfield  {journal}
  {\bibinfo  {journal} {Z.\,Phys.\,C}\ }\textbf {\bibinfo {volume} {220}},\
  \bibinfo {pages} {945} (\bibinfo {year} {2006})}\BibitemShut {NoStop}%
\bibitem [{\citenamefont {Jonsell}(2008)}]{anti:jons08}%
  \BibitemOpen
  \bibfield  {author} {\bibinfo {author} {\bibfnamefont {S.}~\bibnamefont
  {Jonsell}},\ }\bibfield  {title} {\bibinfo {title} {Rearrangement in
  antihydrogen-atom scattering},\ }\href@noop {} {\bibfield  {journal}
  {\bibinfo  {journal} {Nucl.\,Instr.\,Meth.\,B}\ }\textbf {\bibinfo {volume}
  {266}},\ \bibinfo {pages} {369} (\bibinfo {year} {2008})}\BibitemShut
  {NoStop}%
\bibitem [{\citenamefont {Shlyapnikov}\ \emph {et~al.}(1993)\citenamefont
  {Shlyapnikov}, \citenamefont {Walraven},\ and\ \citenamefont
  {Surkov}}]{anti:shly93}%
  \BibitemOpen
  \bibfield  {author} {\bibinfo {author} {\bibfnamefont {G.~V.}\ \bibnamefont
  {Shlyapnikov}}, \bibinfo {author} {\bibfnamefont {J.~T.~M.}\ \bibnamefont
  {Walraven}},\ and\ \bibinfo {author} {\bibfnamefont {E.~L.}\ \bibnamefont
  {Surkov}},\ }\bibfield  {title} {\bibinfo {title} {Antihydrogen at
  sub-{K}elvin temperatures},\ }\href@noop {} {\bibfield  {journal} {\bibinfo
  {journal} {Hyp.\,Int.}\ }\textbf {\bibinfo {volume} {76}},\ \bibinfo {pages}
  {31} (\bibinfo {year} {1993})}\BibitemShut {NoStop}%
\bibitem [{\citenamefont {Fried}\ \emph {et~al.}(1998)\citenamefont {Fried},
  \citenamefont {Killian}, \citenamefont {Willmann}, \citenamefont {Landhuis},
  \citenamefont {Moss}, \citenamefont {Kleppner},\ and\ \citenamefont
  {Greytak}}]{cold:frie98a}%
  \BibitemOpen
  \bibfield  {author} {\bibinfo {author} {\bibfnamefont {D.~G.}\ \bibnamefont
  {Fried}}, \bibinfo {author} {\bibfnamefont {T.~C.}\ \bibnamefont {Killian}},
  \bibinfo {author} {\bibfnamefont {L.}~\bibnamefont {Willmann}}, \bibinfo
  {author} {\bibfnamefont {D.}~\bibnamefont {Landhuis}}, \bibinfo {author}
  {\bibfnamefont {S.~C.}\ \bibnamefont {Moss}}, \bibinfo {author}
  {\bibfnamefont {D.}~\bibnamefont {Kleppner}},\ and\ \bibinfo {author}
  {\bibfnamefont {T.~J.}\ \bibnamefont {Greytak}},\ }\bibfield  {title}
  {\bibinfo {title} {{B}ose-{E}instein condensation of atomic hydrogen},\
  }\href@noop {} {\bibfield  {journal} {\bibinfo  {journal}
  {Phys.\,Rev.\,Lett.}\ }\textbf {\bibinfo {volume} {81}},\ \bibinfo {pages}
  {3811} (\bibinfo {year} {1998})}\BibitemShut {NoStop}%
\bibitem [{\citenamefont {Voronin}\ and\ \citenamefont
  {Carbonell}(1998)}]{anti:voro98}%
  \BibitemOpen
  \bibfield  {author} {\bibinfo {author} {\bibfnamefont {A.~Y.}\ \bibnamefont
  {Voronin}}\ and\ \bibinfo {author} {\bibfnamefont {J.}~\bibnamefont
  {Carbonell}},\ }\bibfield  {title} {\bibinfo {title} {Antiproton--hydrogen
  annihilation at subkelvin temperatures},\ }\href@noop {} {\bibfield
  {journal} {\bibinfo  {journal} {Phys.\,Rev.\,A}\ }\textbf {\bibinfo {volume}
  {57}},\ \bibinfo {pages} {4335} (\bibinfo {year} {1998})}\BibitemShut
  {NoStop}%
\bibitem [{\citenamefont {Froelich}\ \emph {et~al.}(2000)\citenamefont
  {Froelich}, \citenamefont {Jonsell}, \citenamefont {Saenz}, \citenamefont
  {Zygelman},\ and\ \citenamefont {Dalgarno}}]{anti:froe00}%
  \BibitemOpen
  \bibfield  {author} {\bibinfo {author} {\bibfnamefont {P.}~\bibnamefont
  {Froelich}}, \bibinfo {author} {\bibfnamefont {S.}~\bibnamefont {Jonsell}},
  \bibinfo {author} {\bibfnamefont {A.}~\bibnamefont {Saenz}}, \bibinfo
  {author} {\bibfnamefont {B.}~\bibnamefont {Zygelman}},\ and\ \bibinfo
  {author} {\bibfnamefont {A.}~\bibnamefont {Dalgarno}},\ }\bibfield  {title}
  {\bibinfo {title} {Hydrogen-antihydrogen collisions},\ }\href@noop {}
  {\bibfield  {journal} {\bibinfo  {journal} {Phys.\,Rev.\,Lett.}\ }\textbf
  {\bibinfo {volume} {84}},\ \bibinfo {pages} {4577} (\bibinfo {year}
  {2000})}\BibitemShut {NoStop}%
\bibitem [{\citenamefont {Sinha}\ and\ \citenamefont
  {Ghosh}(2000)}]{anti:sinh00}%
  \BibitemOpen
  \bibfield  {author} {\bibinfo {author} {\bibfnamefont {P.~K.}\ \bibnamefont
  {Sinha}}\ and\ \bibinfo {author} {\bibfnamefont {A.~S.}\ \bibnamefont
  {Ghosh}},\ }\bibfield  {title} {\bibinfo {title} {Hydrogen anti-hydrogen
  elastic scattering using fully quantal method},\ }\href@noop {} {\bibfield
  {journal} {\bibinfo  {journal} {Europhys.\,Lett.}\ }\textbf {\bibinfo
  {volume} {49}},\ \bibinfo {pages} {558} (\bibinfo {year} {2000})}\BibitemShut
  {NoStop}%
\bibitem [{\citenamefont {Jonsell}\ \emph {et~al.}(2001)\citenamefont
  {Jonsell}, \citenamefont {Saenz}, \citenamefont {Froelich}, \citenamefont
  {Zygelman},\ and\ \citenamefont {Dalgarno}}]{anti:jons01}%
  \BibitemOpen
  \bibfield  {author} {\bibinfo {author} {\bibfnamefont {S.}~\bibnamefont
  {Jonsell}}, \bibinfo {author} {\bibfnamefont {A.}~\bibnamefont {Saenz}},
  \bibinfo {author} {\bibfnamefont {P.}~\bibnamefont {Froelich}}, \bibinfo
  {author} {\bibfnamefont {B.}~\bibnamefont {Zygelman}},\ and\ \bibinfo
  {author} {\bibfnamefont {A.}~\bibnamefont {Dalgarno}},\ }\bibfield  {title}
  {\bibinfo {title} {Stability of hydrogen--antihydrogen mixtures at low
  energies},\ }\href@noop {} {\bibfield  {journal} {\bibinfo  {journal}
  {Phys.\,Rev.\,A}\ }\textbf {\bibinfo {volume} {64}},\ \bibinfo {pages}
  {052712} (\bibinfo {year} {2001})}\BibitemShut {NoStop}%
\bibitem [{\citenamefont {Armour}\ and\ \citenamefont
  {Chamberlain}(2002)}]{anti:armo02}%
  \BibitemOpen
  \bibfield  {author} {\bibinfo {author} {\bibfnamefont {E.~A.~G.}\
  \bibnamefont {Armour}}\ and\ \bibinfo {author} {\bibfnamefont {C.~W.}\
  \bibnamefont {Chamberlain}},\ }\bibfield  {title} {\bibinfo {title}
  {Calculation of cross sections for very low-energy hydrogen–antihydrogen
  scattering using the {K}ohn variational method},\ }\href@noop {} {\bibfield
  {journal} {\bibinfo  {journal} {J.\,Phys.\,B}\ }\textbf {\bibinfo {volume}
  {35}},\ \bibinfo {pages} {L\,489} (\bibinfo {year} {2002})}\BibitemShut
  {NoStop}%
\bibitem [{\citenamefont {Froelich}\ \emph {et~al.}(2004)\citenamefont
  {Froelich}, \citenamefont {Jonsell}, \citenamefont {Saenz}, \citenamefont
  {Eriksson}, \citenamefont {Zygelman},\ and\ \citenamefont
  {Dalgarno}}]{anti:froe04}%
  \BibitemOpen
  \bibfield  {author} {\bibinfo {author} {\bibfnamefont {P.}~\bibnamefont
  {Froelich}}, \bibinfo {author} {\bibfnamefont {S.}~\bibnamefont {Jonsell}},
  \bibinfo {author} {\bibfnamefont {A.}~\bibnamefont {Saenz}}, \bibinfo
  {author} {\bibfnamefont {S.}~\bibnamefont {Eriksson}}, \bibinfo {author}
  {\bibfnamefont {B.}~\bibnamefont {Zygelman}},\ and\ \bibinfo {author}
  {\bibfnamefont {A.}~\bibnamefont {Dalgarno}},\ }\bibfield  {title} {\bibinfo
  {title} {Leptonic annihilation in hydrogen-antihydrogen collisions},\
  }\href@noop {} {\bibfield  {journal} {\bibinfo  {journal} {Phys.\,Rev.\,A}\
  }\textbf {\bibinfo {volume} {70}},\ \bibinfo {pages} {022509} (\bibinfo
  {year} {2004})}\BibitemShut {NoStop}%
\bibitem [{\citenamefont {Jonsell}\ \emph {et~al.}(2004)\citenamefont
  {Jonsell}, \citenamefont {Saenz}, \citenamefont {Froelich}, \citenamefont
  {Zygelman},\ and\ \citenamefont {Dalgarno}}]{anti:jons04}%
  \BibitemOpen
  \bibfield  {author} {\bibinfo {author} {\bibfnamefont {S.}~\bibnamefont
  {Jonsell}}, \bibinfo {author} {\bibfnamefont {A.}~\bibnamefont {Saenz}},
  \bibinfo {author} {\bibfnamefont {P.}~\bibnamefont {Froelich}}, \bibinfo
  {author} {\bibfnamefont {B.}~\bibnamefont {Zygelman}},\ and\ \bibinfo
  {author} {\bibfnamefont {A.}~\bibnamefont {Dalgarno}},\ }\bibfield  {title}
  {\bibinfo {title} {Hydrogen--antihydrogen scattering in the
  {B}orn-{O}ppenheimer approximation},\ }\href@noop {} {\bibfield  {journal}
  {\bibinfo  {journal} {J.\,Phys.\,B}\ }\textbf {\bibinfo {volume} {37}},\
  \bibinfo {pages} {1195} (\bibinfo {year} {2004})}\BibitemShut {NoStop}%
\bibitem [{\citenamefont {Zygelman}\ \emph {et~al.}(2004)\citenamefont
  {Zygelman}, \citenamefont {Saenz}, \citenamefont {Froelich},\ and\
  \citenamefont {Jonsell}}]{anti:zyge04}%
  \BibitemOpen
  \bibfield  {author} {\bibinfo {author} {\bibfnamefont {B.}~\bibnamefont
  {Zygelman}}, \bibinfo {author} {\bibfnamefont {A.}~\bibnamefont {Saenz}},
  \bibinfo {author} {\bibfnamefont {P.}~\bibnamefont {Froelich}},\ and\
  \bibinfo {author} {\bibfnamefont {S.}~\bibnamefont {Jonsell}},\ }\bibfield
  {title} {\bibinfo {title} {Cold collisions of atomic hydrogen with
  antihydrogen atoms: an optical potential approach},\ }\href@noop {}
  {\bibfield  {journal} {\bibinfo  {journal} {Phys.\,Rev.\,A}\ }\textbf
  {\bibinfo {volume} {69}},\ \bibinfo {pages} {042715} (\bibinfo {year}
  {2004})}\BibitemShut {NoStop}%
\bibitem [{\citenamefont {Armour}\ \emph {et~al.}(2005)\citenamefont {Armour},
  \citenamefont {Liu},\ and\ \citenamefont {Vigier}}]{anti:armo05}%
  \BibitemOpen
  \bibfield  {author} {\bibinfo {author} {\bibfnamefont {E.~A.~G.}\
  \bibnamefont {Armour}}, \bibinfo {author} {\bibfnamefont {Y.}~\bibnamefont
  {Liu}},\ and\ \bibinfo {author} {\bibfnamefont {A.}~\bibnamefont {Vigier}},\
  }\bibfield  {title} {\bibinfo {title} {Inclusion of the strong interaction in
  low-energy hydrogen--antihydrogen scattering using a complex potential},\
  }\href@noop {} {\bibfield  {journal} {\bibinfo  {journal} {J.\,Phys.\,B}\
  }\textbf {\bibinfo {volume} {38}},\ \bibinfo {pages} {L47} (\bibinfo {year}
  {2005})}\BibitemShut {NoStop}%
\bibitem [{\citenamefont {Jonsell}\ \emph {et~al.}(2005)\citenamefont
  {Jonsell}, \citenamefont {Saenz}, \citenamefont {Froelich}, \citenamefont
  {Zygelman},\ and\ \citenamefont {Dalgarno}}]{anti:jons05}%
  \BibitemOpen
  \bibfield  {author} {\bibinfo {author} {\bibfnamefont {S.}~\bibnamefont
  {Jonsell}}, \bibinfo {author} {\bibfnamefont {A.}~\bibnamefont {Saenz}},
  \bibinfo {author} {\bibfnamefont {P.}~\bibnamefont {Froelich}}, \bibinfo
  {author} {\bibfnamefont {B.}~\bibnamefont {Zygelman}},\ and\ \bibinfo
  {author} {\bibfnamefont {A.}~\bibnamefont {Dalgarno}},\ }\bibfield  {title}
  {\bibinfo {title} {Including the strong nuclear force in
  antihydrogen-scattering calculations},\ }\href@noop {} {\bibfield  {journal}
  {\bibinfo  {journal} {Can.\,J.\,Phys.}\ }\textbf {\bibinfo {volume} {83}},\
  \bibinfo {pages} {435} (\bibinfo {year} {2005})}\BibitemShut {NoStop}%
\bibitem [{\citenamefont {Cohen}(2006)}]{anti:cohe06}%
  \BibitemOpen
  \bibfield  {author} {\bibinfo {author} {\bibfnamefont {J.~A.}\ \bibnamefont
  {Cohen}},\ }\bibfield  {title} {\bibinfo {title} {Reactive collisions of
  atomic antihydrogen with {H}, {H}e$^+$ and {H}e},\ }\href@noop {} {\bibfield
  {journal} {\bibinfo  {journal} {J.\,Phys.\,B}\ }\textbf {\bibinfo {volume}
  {39}},\ \bibinfo {pages} {1517} (\bibinfo {year} {2006})}\BibitemShut
  {NoStop}%
\bibitem [{\citenamefont {Jonsell}(2006)}]{anti:jons06}%
  \BibitemOpen
  \bibfield  {author} {\bibinfo {author} {\bibfnamefont {S.}~\bibnamefont
  {Jonsell}},\ }\bibfield  {title} {\bibinfo {title} {Low-temperature
  antihydrogen-atom scattering},\ }\href@noop {} {\bibfield  {journal}
  {\bibinfo  {journal} {Nucl.\,Instr.\,Meth.\,B}\ }\textbf {\bibinfo {volume}
  {247}},\ \bibinfo {pages} {138} (\bibinfo {year} {2006})}\BibitemShut
  {NoStop}%
\bibitem [{\citenamefont {Jonsell}(2018)}]{anti:jons18}%
  \BibitemOpen
  \bibfield  {author} {\bibinfo {author} {\bibfnamefont {S.}~\bibnamefont
  {Jonsell}},\ }\bibfield  {title} {\bibinfo {title} {Collisions involving
  antiprotons and antihydrogen: an overview},\ }\href@noop {} {\bibfield
  {journal} {\bibinfo  {journal} {Phil.\,Trans.\,R.\,Soc.\ A}\ }\textbf
  {\bibinfo {volume} {376}},\ \bibinfo {pages} {20170271} (\bibinfo {year}
  {2018})}\BibitemShut {NoStop}%
\bibitem [{\citenamefont {Zygelman}\ \emph {et~al.}(2001)\citenamefont
  {Zygelman}, \citenamefont {Saenz}, \citenamefont {Froelich}, \citenamefont
  {Jonsell},\ and\ \citenamefont {Dalgarno}}]{anti:zyge01}%
  \BibitemOpen
  \bibfield  {author} {\bibinfo {author} {\bibfnamefont {B.}~\bibnamefont
  {Zygelman}}, \bibinfo {author} {\bibfnamefont {A.}~\bibnamefont {Saenz}},
  \bibinfo {author} {\bibfnamefont {P.}~\bibnamefont {Froelich}}, \bibinfo
  {author} {\bibfnamefont {S.}~\bibnamefont {Jonsell}},\ and\ \bibinfo {author}
  {\bibfnamefont {A.}~\bibnamefont {Dalgarno}},\ }\bibfield  {title} {\bibinfo
  {title} {Radiative association of atomic hydrogen with antihydrogen at
  subkelvin temperatures},\ }\href@noop {} {\bibfield  {journal} {\bibinfo
  {journal} {Phys.\,Rev.\,A}\ }\textbf {\bibinfo {volume} {63}},\ \bibinfo
  {pages} {052722} (\bibinfo {year} {2001})}\BibitemShut {NoStop}%
\bibitem [{\citenamefont {Strasburger}(2002)}]{anti:stra02}%
  \BibitemOpen
  \bibfield  {author} {\bibinfo {author} {\bibfnamefont {K.}~\bibnamefont
  {Strasburger}},\ }\bibfield  {title} {\bibinfo {title} {Accurate
  {B}orn-{O}ppenheimer potential energy curve for the hydrogen-antihydrogen
  system},\ }\href@noop {} {\bibfield  {journal} {\bibinfo  {journal}
  {J.\,Phys.\,B}\ }\textbf {\bibinfo {volume} {35}},\ \bibinfo {pages} {L\,435}
  (\bibinfo {year} {2002})}\BibitemShut {NoStop}%
\bibitem [{\citenamefont {Strasburger}(2004)}]{anti:stra04}%
  \BibitemOpen
  \bibfield  {author} {\bibinfo {author} {\bibfnamefont {K.}~\bibnamefont
  {Strasburger}},\ }\bibfield  {title} {\bibinfo {title} {Hydrogen-antihydrogen
  interaction: spectacular breakdown of the adiabatic approximation},\
  }\href@noop {} {\bibfield  {journal} {\bibinfo  {journal} {J.\,Phys.\,B}\
  }\textbf {\bibinfo {volume} {37}},\ \bibinfo {pages} {4483} (\bibinfo {year}
  {2004})}\BibitemShut {NoStop}%
\bibitem [{\citenamefont {Sinha}\ and\ \citenamefont
  {Ghosh}(2002)}]{anti:sinh02}%
  \BibitemOpen
  \bibfield  {author} {\bibinfo {author} {\bibfnamefont {P.~K.}\ \bibnamefont
  {Sinha}}\ and\ \bibinfo {author} {\bibfnamefont {A.~S.}\ \bibnamefont
  {Ghosh}},\ }\bibfield  {title} {\bibinfo {title} {Capture cross sections at
  low energies for {H}-{$\bar{\textrm{H}}$} scattering},\ }\href@noop {}
  {\bibfield  {journal} {\bibinfo  {journal} {J.\,Phys.\,B}\ }\textbf {\bibinfo
  {volume} {35}},\ \bibinfo {pages} {L281} (\bibinfo {year}
  {2002})}\BibitemShut {NoStop}%
\bibitem [{\citenamefont {Stegeby}\ \emph {et~al.}(2012)\citenamefont
  {Stegeby}, \citenamefont {Piszczatowski}, \citenamefont {Karlsson},
  \citenamefont {Lindh},\ and\ \citenamefont {Froelich}}]{anti:steg12}%
  \BibitemOpen
  \bibfield  {author} {\bibinfo {author} {\bibfnamefont {H.}~\bibnamefont
  {Stegeby}}, \bibinfo {author} {\bibfnamefont {K.}~\bibnamefont
  {Piszczatowski}}, \bibinfo {author} {\bibfnamefont {H.~O.}\ \bibnamefont
  {Karlsson}}, \bibinfo {author} {\bibfnamefont {R.}~\bibnamefont {Lindh}},\
  and\ \bibinfo {author} {\bibfnamefont {P.}~\bibnamefont {Froelich}},\
  }\bibfield  {title} {\bibinfo {title} {Variational calculations for the
  hydrogen–antihydrogen system with a mass–scaled {B}orn–{O}ppenheimer
  potential},\ }\href@noop {} {\bibfield  {journal} {\bibinfo  {journal}
  {Centr. Eur. J. Phys.}\ }\textbf {\bibinfo {volume} {10}},\ \bibinfo {pages}
  {1038} (\bibinfo {year} {2012})}\BibitemShut {NoStop}%
\bibitem [{\citenamefont {Armour}\ and\ \citenamefont
  {Jonsell}(2007)}]{anti:armo07}%
  \BibitemOpen
  \bibfield  {author} {\bibinfo {author} {\bibfnamefont {E.~A.~G.}\
  \bibnamefont {Armour}}\ and\ \bibinfo {author} {\bibfnamefont
  {S.}~\bibnamefont {Jonsell}},\ }\bibfield  {title} {\bibinfo {title} {On the
  evaluation of transition matrices for rearrangement in atom–antiatom
  scattering},\ }\href@noop {} {\bibfield  {journal} {\bibinfo  {journal}
  {J.\,Phys.\,A}\ }\textbf {\bibinfo {volume} {40}},\ \bibinfo {pages} {701}
  (\bibinfo {year} {2007})}\BibitemShut {NoStop}%
\bibitem [{\citenamefont {Voronin}\ and\ \citenamefont
  {Froelich}(2008)}]{anti:voro08}%
  \BibitemOpen
  \bibfield  {author} {\bibinfo {author} {\bibfnamefont {A.~Y.}\ \bibnamefont
  {Voronin}}\ and\ \bibinfo {author} {\bibfnamefont {P.}~\bibnamefont
  {Froelich}},\ }\bibfield  {title} {\bibinfo {title} {Resonant phenomena in
  antihydrogen-hydrogen scattering},\ }\href@noop {} {\bibfield  {journal}
  {\bibinfo  {journal} {Phys.\,Rev.\,A}\ }\textbf {\bibinfo {volume} {77}},\
  \bibinfo {pages} {022505} (\bibinfo {year} {2008})}\BibitemShut {NoStop}%
\bibitem [{\citenamefont {Piszczatowski}\ \emph
  {et~al.}(2014{\natexlab{a}})\citenamefont {Piszczatowski}, \citenamefont
  {Voronin},\ and\ \citenamefont {Froelich}}]{anti:pisz14a}%
  \BibitemOpen
  \bibfield  {author} {\bibinfo {author} {\bibfnamefont {K.}~\bibnamefont
  {Piszczatowski}}, \bibinfo {author} {\bibfnamefont {A.}~\bibnamefont
  {Voronin}},\ and\ \bibinfo {author} {\bibfnamefont {P.}~\bibnamefont
  {Froelich}},\ }\bibfield  {title} {\bibinfo {title} {Nonadiabatic treatment
  of hydrogen-antihydrogen collisions},\ }\href@noop {} {\bibfield  {journal}
  {\bibinfo  {journal} {Phys.\,Rev.\,A}\ }\textbf {\bibinfo {volume} {89}},\
  \bibinfo {pages} {062703} (\bibinfo {year} {2014}{\natexlab{a}})}\BibitemShut
  {NoStop}%
\bibitem [{\citenamefont {Piszczatowski}\ \emph
  {et~al.}(2014{\natexlab{b}})\citenamefont {Piszczatowski}, \citenamefont
  {Voronin},\ and\ \citenamefont {Froelich}}]{anti:pisz14b}%
  \BibitemOpen
  \bibfield  {author} {\bibinfo {author} {\bibfnamefont {K.}~\bibnamefont
  {Piszczatowski}}, \bibinfo {author} {\bibfnamefont {A.}~\bibnamefont
  {Voronin}},\ and\ \bibinfo {author} {\bibfnamefont {P.}~\bibnamefont
  {Froelich}},\ }\bibfield  {title} {\bibinfo {title} {Four-body calculations
  of elastic scattering in {H}-{$\bar{\rm H}$} collisions},\ }\href@noop {}
  {\bibfield  {journal} {\bibinfo  {journal} {Hyp.\,Int.}\ }\textbf {\bibinfo
  {volume} {228}},\ \bibinfo {pages} {85} (\bibinfo {year}
  {2014}{\natexlab{b}})}\BibitemShut {NoStop}%
\bibitem [{\citenamefont {Stegeby}\ and\ \citenamefont
  {Piszczatowski}(2016)}]{anti:steg16}%
  \BibitemOpen
  \bibfield  {author} {\bibinfo {author} {\bibfnamefont {H.}~\bibnamefont
  {Stegeby}}\ and\ \bibinfo {author} {\bibfnamefont {K.}~\bibnamefont
  {Piszczatowski}},\ }\bibfield  {title} {\bibinfo {title} {Resonance states in
  the hydrogen–antihydrogen system from a nonadiabatic treatment},\
  }\href@noop {} {\bibfield  {journal} {\bibinfo  {journal} {J.\,Phys.\,B}\
  }\textbf {\bibinfo {volume} {49}},\ \bibinfo {pages} {014002} (\bibinfo
  {year} {2016})}\BibitemShut {NoStop}%
\bibitem [{\citenamefont {Yamashita}\ and\ \citenamefont
  {Kino}(2018)}]{anti:yama18b}%
  \BibitemOpen
  \bibfield  {author} {\bibinfo {author} {\bibfnamefont {T.}~\bibnamefont
  {Yamashita}}\ and\ \bibinfo {author} {\bibfnamefont {Y.}~\bibnamefont
  {Kino}},\ }\bibfield  {title} {\bibinfo {title} {Coupled channel study of
  antihydrogen-hydrogen molecular resonance state},\ }\href@noop {} {\bibfield
  {journal} {\bibinfo  {journal} {JJAP Conference Proceedings}\ }\textbf
  {\bibinfo {volume} {7}},\ \bibinfo {pages} {011004} (\bibinfo {year}
  {2018})}\BibitemShut {NoStop}%
\bibitem [{\citenamefont {Hiyama}\ \emph {et~al.}(2003)\citenamefont {Hiyama},
  \citenamefont {Kino},\ and\ \citenamefont {Kamimura}}]{anti:hiya03}%
  \BibitemOpen
  \bibfield  {author} {\bibinfo {author} {\bibfnamefont {E.}~\bibnamefont
  {Hiyama}}, \bibinfo {author} {\bibfnamefont {Y.}~\bibnamefont {Kino}},\ and\
  \bibinfo {author} {\bibfnamefont {M.}~\bibnamefont {Kamimura}},\ }\bibfield
  {title} {\bibinfo {title} {Gaussian expansion method for few-body systems},\
  }\href@noop {} {\bibfield  {journal} {\bibinfo  {journal}
  {Prog.\,Part.\,Nucl.\,Phys.}\ }\textbf {\bibinfo {volume} {51}},\ \bibinfo
  {pages} {223} (\bibinfo {year} {2003})}\BibitemShut {NoStop}%
\bibitem [{\citenamefont {Labzowsky}\ \emph {et~al.}(2005)\citenamefont
  {Labzowsky}, \citenamefont {Sharipov}, \citenamefont {Prozorov},
  \citenamefont {Plunien},\ and\ \citenamefont {Soff}}]{anti:labz05}%
  \BibitemOpen
  \bibfield  {author} {\bibinfo {author} {\bibfnamefont {L.}~\bibnamefont
  {Labzowsky}}, \bibinfo {author} {\bibfnamefont {V.}~\bibnamefont {Sharipov}},
  \bibinfo {author} {\bibfnamefont {A.}~\bibnamefont {Prozorov}}, \bibinfo
  {author} {\bibfnamefont {G.}~\bibnamefont {Plunien}},\ and\ \bibinfo {author}
  {\bibfnamefont {G.}~\bibnamefont {Soff}},\ }\bibfield  {title} {\bibinfo
  {title} {Decay channels and decay rates for the hydrogen-antihydrogen
  quasimolecule},\ }\href {https://doi.org/10.1103/PhysRevA.72.022513}
  {\bibfield  {journal} {\bibinfo  {journal} {Phys.\,Rev.\,A}\ }\textbf
  {\bibinfo {volume} {72}},\ \bibinfo {pages} {022513} (\bibinfo {year}
  {2005})}\BibitemShut {NoStop}%
\bibitem [{\citenamefont {Ferr\'on}\ \emph {et~al.}(2008)\citenamefont
  {Ferr\'on}, \citenamefont {Serra},\ and\ \citenamefont {Kais}}]{anti:ferr08}%
  \BibitemOpen
  \bibfield  {author} {\bibinfo {author} {\bibfnamefont {A.}~\bibnamefont
  {Ferr\'on}}, \bibinfo {author} {\bibfnamefont {P.}~\bibnamefont {Serra}},\
  and\ \bibinfo {author} {\bibfnamefont {S.}~\bibnamefont {Kais}},\ }\bibfield
  {title} {\bibinfo {title} {Stability conditions for
  hydrogen-antihydrogen--like quasimolecules},\ }\href
  {https://doi.org/10.1103/PhysRevA.77.052505} {\bibfield  {journal} {\bibinfo
  {journal} {Phys.\,Rev.\,A}\ }\textbf {\bibinfo {volume} {77}},\ \bibinfo
  {pages} {052505} (\bibinfo {year} {2008})}\BibitemShut {NoStop}%
\bibitem [{\citenamefont {Sharipov}\ \emph
  {et~al.}(2006{\natexlab{a}})\citenamefont {Sharipov}, \citenamefont
  {Labzowsky},\ and\ \citenamefont {Plunien}}]{anti:shar06}%
  \BibitemOpen
  \bibfield  {author} {\bibinfo {author} {\bibfnamefont {V.}~\bibnamefont
  {Sharipov}}, \bibinfo {author} {\bibfnamefont {L.}~\bibnamefont
  {Labzowsky}},\ and\ \bibinfo {author} {\bibfnamefont {G.}~\bibnamefont
  {Plunien}},\ }\bibfield  {title} {\bibinfo {title} {Potential energy curves
  for excited states of the hydrogen--antihydrogen system},\ }\href@noop {}
  {\bibfield  {journal} {\bibinfo  {journal} {Phys.\,Rev.\,Lett.}\ }\textbf
  {\bibinfo {volume} {97}},\ \bibinfo {pages} {103005} (\bibinfo {year}
  {2006}{\natexlab{a}})}\BibitemShut {NoStop}%
\bibitem [{\citenamefont {Sharipov}\ \emph
  {et~al.}(2006{\natexlab{b}})\citenamefont {Sharipov}, \citenamefont
  {Labzowsky},\ and\ \citenamefont {Plunien}}]{anti:shar06a}%
  \BibitemOpen
  \bibfield  {author} {\bibinfo {author} {\bibfnamefont {V.}~\bibnamefont
  {Sharipov}}, \bibinfo {author} {\bibfnamefont {L.}~\bibnamefont
  {Labzowsky}},\ and\ \bibinfo {author} {\bibfnamefont {G.}~\bibnamefont
  {Plunien}},\ }\bibfield  {title} {\bibinfo {title} {Rydberg states of the
  hydrogen-antihydrogen quasimolecule},\ }\href@noop {} {\bibfield  {journal}
  {\bibinfo  {journal} {Phys.\,Rev.\,A}\ }\textbf {\bibinfo {volume} {73}},\
  \bibinfo {pages} {052503} (\bibinfo {year} {2006}{\natexlab{b}})}\BibitemShut
  {NoStop}%
\bibitem [{\citenamefont {Pachucki}\ \emph {et~al.}(2016)\citenamefont
  {Pachucki}, \citenamefont {Zientkiewizc},\ and\ \citenamefont
  {Yerokhin}}]{dia:pach16}%
  \BibitemOpen
  \bibfield  {author} {\bibinfo {author} {\bibfnamefont {K.}~\bibnamefont
  {Pachucki}}, \bibinfo {author} {\bibfnamefont {M.}~\bibnamefont
  {Zientkiewizc}},\ and\ \bibinfo {author} {\bibfnamefont {V.}~\bibnamefont
  {Yerokhin}},\ }\bibfield  {title} {\bibinfo {title} {{$\mathrm{H2SOLV:}$
  Fortran solver for diatomic molecules in explicitly correlated exponential
  basis}},\ }\href@noop {} {\bibfield  {journal} {\bibinfo  {journal} {Computer
  Physics Communications}\ }\textbf {\bibinfo {volume} {208}},\ \bibinfo
  {pages} {162} (\bibinfo {year} {2016})}\BibitemShut {NoStop}%
\bibitem [{Note1()}]{Note1}%
  \BibitemOpen
  \bibinfo {note} {The definition of the base parameters of Pachucki et al.\
  differs slightly from the original one introduced by W.~Ko\l {}os and
  L.~Wolniewicz\ who used the symbols $\protect \{\alpha ,\bar {\alpha },\beta
  ,\bar {\beta }\protect \}$. They are related by $uR = \alpha $, $wR = \bar
  {\alpha }$, $yR = -\beta $, $xR=-\bar {\beta }$. The motivation for dividing
  the Ko\l {}os-Wolniewicz\ parameters by $R$ is that in this case the
  exponential damping becomes independent of the inter-hadronic separation,
  since the implicit dependence contained in the prolate-spheroidal coordinates
  $\xi $ and $\eta $ is canceled.}\BibitemShut {Stop}%
\bibitem [{\citenamefont {Bailey}()}]{gen:MPFUN2015}%
  \BibitemOpen
  \bibfield  {author} {\bibinfo {author} {\bibfnamefont {D.~H.}\ \bibnamefont
  {Bailey}},\ }\href {https://www.davidhbailey.com/dhbsoftware/} {\bibinfo
  {title} {{MPFUN2015} a thread-safe arbitrary precision package}},\ \bibinfo
  {note} {\url{https://www.davidhbailey.com/dhbsoftware/}}\BibitemShut
  {NoStop}%
\bibitem [{Note2()}]{Note2}%
  \BibitemOpen
  \bibinfo {note} {Convergence plots for additional $R$ values and also for the
  10 lowest lying Q-odd states are given in Appendix \ref
  {sec:BO_curves}.}\BibitemShut {Stop}%
\bibitem [{\citenamefont {Wolniewicz}(1993)}]{dia:woln93}%
  \BibitemOpen
  \bibfield  {author} {\bibinfo {author} {\bibfnamefont {L.}~\bibnamefont
  {Wolniewicz}},\ }\bibfield  {title} {\bibinfo {title} {{Relativistic energies
  of the ground state of the hydrogen molecule}},\ }\href@noop {} {\bibfield
  {journal} {\bibinfo  {journal} {J.\,Chem.\,Phys.}\ }\textbf {\bibinfo
  {volume} {99}},\ \bibinfo {pages} {1851} (\bibinfo {year}
  {1993})}\BibitemShut {NoStop}%
\bibitem [{\citenamefont {Staszewska}\ and\ \citenamefont
  {Wolniewicz}(1999)}]{dia:stas99}%
  \BibitemOpen
  \bibfield  {author} {\bibinfo {author} {\bibfnamefont {G.}~\bibnamefont
  {Staszewska}}\ and\ \bibinfo {author} {\bibfnamefont {L.}~\bibnamefont
  {Wolniewicz}},\ }\bibfield  {title} {\bibinfo {title} {{Transition moments
  among $^3\Sigma$ and $^3\Pi$ states of the $\rm{H}_2$ molecule}},\
  }\href@noop {} {\bibfield  {journal} {\bibinfo  {journal}
  {J.\,Mol.\,Spectrosc.}\ }\textbf {\bibinfo {volume} {198}},\ \bibinfo {pages}
  {416} (\bibinfo {year} {1999})}\BibitemShut {NoStop}%
\bibitem [{\citenamefont {Staszewska}\ and\ \citenamefont
  {Wolniewicz}(2002)}]{dia:stas02}%
  \BibitemOpen
  \bibfield  {author} {\bibinfo {author} {\bibfnamefont {G.}~\bibnamefont
  {Staszewska}}\ and\ \bibinfo {author} {\bibfnamefont {L.}~\bibnamefont
  {Wolniewicz}},\ }\bibfield  {title} {\bibinfo {title} {{Adiabatic energies of
  excited $^1{\Sigma}_u$ states of the hydrogen molecule}},\ }\href@noop {}
  {\bibfield  {journal} {\bibinfo  {journal} {J.\,Mol.\,Spectrosc.}\ }\textbf
  {\bibinfo {volume} {212}},\ \bibinfo {pages} {208} (\bibinfo {year}
  {2002})}\BibitemShut {NoStop}%
\bibitem [{\citenamefont {Bubin}\ and\ \citenamefont
  {Adamowicz}(2020)}]{gen:bubi20}%
  \BibitemOpen
  \bibfield  {author} {\bibinfo {author} {\bibfnamefont {S.}~\bibnamefont
  {Bubin}}\ and\ \bibinfo {author} {\bibfnamefont {L.}~\bibnamefont
  {Adamowicz}},\ }\bibfield  {title} {\bibinfo {title} {{Computer program
  ATOM-MOL-nonBO for performing calculations of ground and excited states of
  atoms and molecules without assuming the Born–Oppenheimer approximation
  using all-particle complex explicitly correlated Gaussian functions}},\
  }\href@noop {} {\bibfield  {journal} {\bibinfo  {journal} {J.\,Chem.\,Phys.}\
  }\textbf {\bibinfo {volume} {152}} (\bibinfo {year} {2020})}\BibitemShut
  {NoStop}%
\end{thebibliography}
%%%
%

\end{document}